\documentclass[article,accept,moreauthors,galaxies,pdftex,10pt,a4paper]{mdpi} 
\usepackage{graphicx}
\usepackage{tikz}
\firstpage{1} 
\articlenumber{x}
\doinum{10.3390/------}
\pubvolume{xx}
\pubyear{2018}
\copyrightyear{2018}
\externaleditor{Academic Editor: name}
\history{Received: date; Accepted: date; Published: date}

 \theoremstyle{mdpi}
 \newcounter{thm}
 \newcounter{ex}
 \newcounter{re}
 \theoremstyle{mdpidefinition}

\Title{Strongly Magnetized Sources: QED and X-ray Polarization}

\Author{Jeremy Heyl and Ilaria Caiazzo}
\AuthorNames{Jeremy Heyl and Ilaria Caiazzo}

\address{%
Department of Physics and Astronomy, 6224 Agricultural Road, University of British Columbia, Vancouver BC V6T 1Z1; heyl@phas.ubc.ca, ilariacaiazzo@phas.ubc.ca}

\corres{Correspondence: heyl@phas.ubc.ca; Tel.: +1-604-822-0995}

\abstract{Radiative corrections of quantum electrodynamics cause a vacuum threaded by magnetic field to be birefringent. This means that radiation of different polarizations travels at different speeds. Even in the strong magnetic fields of astrophysical sources the difference in speed is small; however, it has profound consequences for the extent of polarization expected from strongly magnetized sources. We demonstrate how the birefringence arises from first principles, show how birefringence affects the polarization state of radiation and present recent calculations for the expected polarization from magnetars and X-ray pulsars.}

\keyword{quantum electrodynamics: radiative corrections; magnetic fields; neutron stars; X-ray polarization}

\def\tilde{\mathaccent"365}			
\def\ie{{\it i.e.~}}

\newcommand{\be}{\begin{equation}}
\newcommand{\ee}{\end{equation}}
\newcommand{\ba}{\begin{eqnarray}}
\newcommand{\ea}{\end{eqnarray}}
\def\p{\partial}
\def\d{{\rm d}}

\def\dd#1#2{\frac{\d #1}{\d #2}}
\def\pp#1#2{\frac{\p #1}{\p #2}}
\def\ff#1#2{\frac{\delta #1}{\delta #2}}
\def\eqref#1{Eq.~\ref{eq:#1}}
\newcommand{\bfi}{{\vec B}}
\newcommand{\efi}{{\vec E}}
\newcommand{\lag}{{\cal L}}
\def\D{{\cal D}}
\def\L{{\cal L}}
\def\4d#1{\d^4\!#1}

\def\bpsi{{\bar \psi}}
\def\Tr{\mathrm{Tr}}
\def\tr{\mathrm{tr}}
\def\Det{\mathrm{Det}}

\begin{document}




\section{Introduction}

The second prediction of quantum electrodynamics (QED) was the birefringence of the vacuum \cite{Heis36,Weis36}, in particular that light travels through a magnetic field at different speeds depending on its polarization. This came only eight years after the formulation of relativistic quantum mechanics \cite{1928RSPSA.117..610D} and five years after the first prediction of QED, the existence of the positron \cite{1931RSPSA.133...60D}.  Of course, the positron was discovered soon after \cite{1933PhRv...43..491A}, but vacuum birefringence has not yet been observed definitively, although recently astronomers have discovered strong evidence for it \cite{2017MNRAS.465..492M}.  During the next fifteen years, as QED was formalized and place on a rigorous foundation, the prediction of vacuum birefringence was also made more rigorous \cite{Schw51}.

Why has it taken so long to observe directly the second prediction of QED?  The primary reason is that, for magnetic field strengths achieved in the laboratory, the effect is incredibly small. At one Tesla, the difference in the index of refraction between the two modes is $4\times 10^{-24}$. Despite the incredible smallness of the effect, terrestrial experiments are reaching closer and closer to measuring it \cite{2012IJMPA..2760017Z,2017RScI...88l3114H} in strong magnetic fields.  The development of more powerful lasers promises to probe the closely related effect of photon-photon  scattering \cite{2016PhRvA..94f2102K,2017JPhCS.869a2015H}.  Even in the magnetic fields of the most strongly magnetized objects in the Universe, the magnetars, the difference in the index of refraction is only a few percent; however, the vacuum birefringence can increase the observed extent of polarization in the X-rays by a factor of ten even for more weakly magnetized objects.

Although the difference in the index of refraction is small between the modes, this difference means that, for radiation of sufficiently high frequency, the polarization states will not mix as they travel through the magnetosphere of the source \cite{1987PASJ...39..781K}. If the emission originates from a region over which the local magnetic field varies in direction, the decoupling of the polarization modes by vacuum polarization makes the final polarization fraction up to ten times larger than without it. This was first discovered by \cite{Heyl01qed} and fueled a resurgence in the interest of using astrophysics to test QED.

Rather than provide a detailed review of all of the possible QED effects, we will focus on motivating why QED is important for X-ray polarization from magnetized objects. We will start from first principles, leading to some recent results for magnetars and X-ray pulsars that can be probed with instruments currently under development \cite{2016SPIE.9905E..17W,2016arXiv160708823Z,2016SPIE.9905E..15S}. We will leave out other interesting phenomena, such as the role that the competition between vacuum and plasma birefringence plays in the formation of spectral lines \cite{2003MNRAS.338..233H}. In particular, we will derive the effective Lagrangian of QED from the classical Lagrangian of the electromagnetic field coupled to a Dirac field using modern functional techniques (\S~\ref{sec:effaction_derivation}), we will calculate the index of refraction (\S~\ref{sec:index}) and how it affects the propagation of polarization radiation (\S~\ref{sec:propagation}), and finally we will present recent results for polarized emission from magnetars and X-ray pulsars (\S~\ref{sec:results}).  

\section{Theoretical Treatment : Effective Action and Vacuum Birefringence}

If an external field breaks the symmetry of the vacuum, we must develop new tools to account for its presence.  The Lagrangian of QED is
\begin{equation}
\L = \bpsi (i \gamma^\mu \pp{}{x^\mu}+ e \gamma^\mu A_\mu - m) \psi 
- \frac{1}{4} F_{\mu\nu} F^{\mu\nu} 
\label{eq:1}
\end{equation}
where the interaction between the electromagnetic fields and the external field is given by the Feynman rule
\begin{equation}
- i e \gamma^\mu {\tilde{A}}^0_\mu ({\bf q}) \;.
\label{eq:2}
\end{equation}
This rule must be taken into account in all fermion propagators, including internal lines such as in the vacuum polarization and in photon splitting processes. The symbols $\gamma^\mu$ are the Dirac matrices that span the spinorial components of the fermion fields. 

If the external field is sufficiently weak, the interaction with the field may be treated perturbatively as a series of discrete interactions. On the other hand, at a field strength of $B_\mathrm{\scriptsize QED}= m^2 c^3 / (e \hbar) =  4.4 \times 10^{13}$~G, the gyration energy of an electron or, equivalently, the potential energy drop across its Compton wavelength, is equal to its rest mass.  For this reason, when the field exceeds a critical value of approximately $B_\mathrm{\scriptsize QED}/2$, this series fails to converge.  Essentially, each term in the sum of diagrams is equally large in this limit. 

In the next section, we derive the effective action of a general field configuration to one-loop order using the QED Lagrangian (Eq.~\ref{eq:1}) and techniques from statistical mechanics. The key results for X-ray polarization are the index of refraction (\S~\ref{sec:index}) and how polarization changes as radiation propagates through an inhomogeneous birefringent medium (\S~\ref{sec:propagation}). 

\subsection{Effective Action : Formal Derivation}
\label{sec:effaction_derivation}
The connections between the theory of quantum and statistical fields are manifold. Our derivation of the effective action and Lagrangian will exploit these connections. The final results that we present here are well known in the specialized literature for quantum field theory in strong fields \cite[e.g][]{Ditt85,Ditt00} and older introductory texts \cite[e.g.][]{Itzy80,Bere82}, but they are typically absent from recent introductory texts \cite[e.g.][]{Mand93,Pesk95}.  We present a derivation of the effective action using functional techniques familiar from modern treatments of quantum field theory \cite[e.g.][]{Pesk95} and statistical mechanics.

First, the partition function consists of the sum of the quantum phases (or statistical weights) of each possible state of the system (the analysis in this section draws on \S \S 11.3-11.4 of \cite{Pesk95}),
\begin{equation} 
Z[J^\mu, {\bar \eta}, \eta] = \exp \left ( -\frac{i}{\hbar} E [J^\mu,
{\bar \eta}, \eta] \right ) = \int \D A_\mu \D \psi \D \bpsi \exp
\frac{i}{\hbar} \int \4dx \left ( \L + J^\mu A_\mu + {\bar \eta} \psi
+ \bpsi \eta \right ).  
\label{eq:3}
\end{equation}
where $J^\mu$ and $A_\mu$ are the electromagnetic current and vector potential respectively,  $\eta$ and $\bar \eta$  are the fermionic currents and $\psi$ and $\bar \psi$ are the fermionic fields. The variables $\eta, \bar \eta, \psi$ and $\bar \psi$ are anti-commuting Grassman numbers. This means that $\eta \psi = -\psi \eta$ and $\eta^2 = 0$. Grassman numbers behave differently from commuting numbers under integration and differentiation as well. Brackets are used to denote functionals (integrals of the fields over the entire spacetime) while parentheses indicate functions.

Each field configuration receives a phase proportional to the integral of the Lagrangian over spacetime, \ie the action.  The constant of proportionality is $i/\hbar$; in statistical physics, the constant of proportionality is $-1/(kT)$, and the states are weighted by energy and not action. Drawing the analogy further, the functional $E[J^\mu, {\bar \eta}, \eta]$, would correspond to the Helmholtz free energy $F(T,V,N)$ in statistical mechanics; it is the vacuum energy as a function of the external sources $J^\mu, {\bar \eta}$ and $\eta$. The partial derivative $F(T,V,N)$ with respect to the volume of the system yields the thermodynamic conjugate to $V$, the pressure. For the quantum field, the result proceeds similarly,
\begin{eqnarray}
\ff{E[J^\mu, {\bar \eta}, \eta]}{J^\mu(x)} &=& i \hbar \ff{}{J^\mu(x)} \ln Z =
- \frac { \int \D A_\mu \D \psi \D
\bpsi A_\mu(x) \exp \frac{i}{\hbar} \int \4dx \left ( \L + J^\mu A_\mu + {\bar \eta} \psi +
\bpsi \eta \right ) }{\int \D A_\mu \D \psi \D
\bpsi \exp \frac{i}{\hbar} \int \4dx \left ( \L + J^\mu A_\mu + {\bar \eta} \psi +
\bpsi \eta \right )} \\
&=& - \left < \Omega | A_\mu(x) | \Omega \right > \equiv - A^0_\mu(x).
\label{eq:5}
\end{eqnarray} 
where we use the symbol $\delta$ to denote a functional derivative and where $\left | \Omega \right >$ denotes the vacuum state. The functional derivative of $E[~]$ with respect to one of the currents yields the classical field, \ie the vacuum expectation value of the corresponding field, which we denote as $A_\mu^0(x)$.

Generally, when one considers the properties of the magnetized vacuum, it is the fields that are specified, not the currents. The effective action is related to $E[~]$ through a Legendre transformation, just as the Gibbs free energy $G$ is related to $F$, \ie
\begin{equation}
G = F - V \left . \pp{F}{V} \right |_T = F + P V
\label{eq:6}
\end{equation}
Analogously, the effective action is
\begin{eqnarray}
\Gamma[ A^0_\mu, \bpsi^0, \psi^0 ] = - E[J^\mu, {\bar \eta}, \eta] - \int \4dy
\left (  
J^\mu(y) A^0_\mu(y)  + {\bar \eta}(y) \psi^0(y)  +  \bpsi^0(y) \eta(y) 
  \right ).
\label{eq:7}
\end{eqnarray}
The functional derivative of the effective action, $\Gamma [ ~ ]$, with respect to one of the classical fields yields the distribution of the corresponding current. Using the analogy with thermodynamics, the effective action is the vacuum energy with the distribution of the fields fixed.

\subsubsection{Functional integration}

Computing the effective action begins with the expression for $Z[~]$, the partition function, specifically by expanding the classical action with currents about the values of the classical fields, 
\begin{eqnarray}
\int \4dx \left ( \L + J^\mu A_\mu + {\bar \eta} \psi + \bpsi \eta \right ) &=&
\int \4dx \left ( \L [ A^0_\mu, \bpsi^0, \psi^0 ] + J^\mu A^0_\mu 
	+ {\bar \eta} \psi^0 + \bpsi^0 \eta  \right )
 + \nonumber \\ 
& &  \int \4dx \left [
\Delta A_\mu(x)  \left ( \ff{\L}{A_\mu} - J^\mu \right ) + 
\Delta \bpsi(x) \left ( \ff{\L}{\bpsi} - {\bar \eta} \right ) +
 \left ( \ff{\L}{\psi} - \eta \right ) \Delta \psi(x)
\right ] + \nonumber \\
& & \frac{1}{2} 
\int \4dx \4dy  \Biggr [ ( \Delta A_\mu(x) ) ( \Delta A_\nu(y) )
\frac{\delta^2 \L}{\delta A_\mu(x) \delta A_\nu(y) } +  
\nonumber \\
& & ~~~
( \Delta \bpsi(x) )
\frac{\delta^2 \L}{\delta \bpsi(x) \delta \psi(y) }  ( \Delta \psi(y) ) +
\nonumber \\
& & ~~~ ( \Delta A_\mu(x) ) 
\frac{\delta^2 \L}{\delta A_\mu(x) \delta \psi(y) }  ( \Delta \psi(y) ) 
\nonumber \\
& & ~~~
+ 
( \Delta \bpsi(x) )
\frac{\delta^2 \L}{\delta \bpsi(x) \delta A_\mu(y) }  ( \Delta A_\mu(y) )
\Biggr ] +
\mathrm{Higher~Order~Terms}
\label{eq:8}
\end{eqnarray}
where $\Delta A_\mu(x) = A_\mu(x) - A_\mu^0(x)$, the difference between the electromagnetic field including the quantum fluctuations and the classical electromagnetic field, and similarly for the other fields.
Since the functional
 derivatives will be evaluated at
$\psi^0(y)=\bpsi^0(y)=0$, the last two terms vanish.   Furthermore, second derivatives with respect to the same Grassman field also vanish because of the anti-commutative nature of the fields.  Let's evaluate for an example the term
\begin{eqnarray}
\ff{\L}{A_\mu} - J^\mu &=& \ff{}{A_\mu} \left ( \bpsi (i \gamma^\mu \pp{}{x^\mu}+ e \gamma^\mu A_\mu - m) \psi 
- \frac{1}{4} F_{\mu\nu} F^{\mu\nu} \right ) - J^\mu \\
&=& \bpsi e \gamma^\mu \psi - J^\mu - \frac{1}{4} \ff{}{A_\mu} \left(
A_{\mu,\nu} A^{\mu,\nu} - A_{\nu,\mu} A^{\mu,\nu} - A_{\mu,\nu}
A^{\nu,\mu} + A_{\nu,\mu} A^{\nu,\mu} \right ) \\
&=& \bpsi e \gamma^\mu \psi - J^\mu - \frac{1}{4}  
\ff{}{A_\mu} \left(
2 A_{\mu,\nu} A^{\mu,\nu} - 2 A_{\mu,\nu} A^{\nu,\mu} \right ) \\
&=& 
\bpsi e \gamma^\mu \psi - J^\mu -  \frac{\stackrel{\leftarrow}{\partial}}{\partial x^\nu} A^{\mu,\nu} +
\frac{\stackrel{\leftarrow}{\partial}}{\partial x^\nu} A^{\nu,\mu}
\label{eq:9}
\end{eqnarray}
The $\stackrel{\leftarrow}{\partial}$ notation indicates that the resulting functional derivative is an operator that differentiates something to the left.  Because the function derivative lives within a integral over all of spacetime, we can use integration by parts to simplify the result further if we assume that the boundary terms vanish:
\begin{equation}
\ff{\L}{A_\mu} - J^\mu =
\bpsi e \gamma^\mu \psi - J^\mu + \pp{}{x^\nu} \left (
A^{\mu,\nu}- A^{\nu,\mu} \right ) =
\bpsi e \gamma^\mu \psi - J^\mu + \pp{}{x^\nu} F^{\mu\nu}
\label{eq:10}
\end{equation}
We see that the first-order derivatives vanish when the fields satisfy the field equations.

Although we do not have an explicit relationship that connects the currents to the classical fields that they generate, we will impose that the currents $J^\mu$, ${\bar \eta}$ and $\eta$ along with the classical fields $A^0_\mu$, $\bpsi^0$ and $\psi^0$ satisfy the field equations and evaluate all of the functional derivatives at the values of the classical fields, so the first order terms in the expansion vanish and the vacuum energy $E[~]$ to lowest order is a Gaussian functional integral,
\begin{eqnarray}
E[J^\mu, {\bar \eta}, \eta] &=& 
-\int \4dx \left ( - \frac{1}{4} F^0_{\mu\nu} F^{0,\mu\nu}  + 
J^\mu A^0_\mu  \right )  + \nonumber \\
& & ~~~~
i \hbar \ln
\int \D A_\mu \D \bpsi \D \psi \exp \frac{i}{2\hbar}
\int \4dx \4dy  \biggr [ ( \Delta A_\mu(x) ) ( \Delta A_\nu(y) )
\frac{\delta^2 \L}{\delta A_\mu(x) \delta A_\nu(y) } \nonumber
\\
& & ~~~~~~~~~~~~~~~~~~~~~~~~~~~~~~~~~~~~~~~~~~~~~~~~~~~~~~~ + 
( \Delta \bpsi(x) )
\frac{\delta^2 \L}{\delta \bpsi(x) \delta \psi(y) }  ( \Delta \psi(y) ) 
\biggr ] \\
&=& -\int \4dx \left ( - \frac{1}{4} F^0_{\mu\nu} F^{0,\mu\nu}  + 
J^\mu A^0_\mu  \right )  
\nonumber \\
& & ~~~~ 
-\frac{i\hbar}{2} \ln \Det \left [
\frac{\delta^2 \L}{\delta A_\mu(x) \delta A_\nu(y) } \right ] 
+i\hbar \ln \Det \left [ 
\frac{\delta^2 \L}{\delta \bpsi(x) \delta \psi(y) } \right ] 
\nonumber \\
& & ~~~~ +
\mathrm{Constant~Terms}
\label{eq:11}
\end{eqnarray}
We integrated over all of the possible field configurations by assuming that a particular field configuration is a point in an infinite dimensional space and by looking at the functional derivatives as infinite dimensional matrices.  Let us look at the following term in detail to see how this works:
\begin{eqnarray}
E_{A_\mu} &=& i \hbar \ln
\int \D A_\mu \exp \frac{i}{2\hbar}
\int \4dx \4dy  \left [ ( \Delta A_\mu(x) ) ( \Delta A_\nu(y) )
\frac{\delta^2 \L}{\delta A_\mu(x) \delta A_\nu(y) } \right ] \\
&=& \int \prod \left ( \d A_\mu(x_i) \right ) \exp \frac{i}{2\hbar} \sum_{i,j}  ( \Delta A_\mu(x_i) ) ( \Delta A_\nu(y_j) )
\left [\frac{\delta^2 \L}{\delta A_\mu(x_i) \delta A_\nu(y_j) } \right ] \\
&=& \int \prod \left ( \d A_\mu(x_i) \right ) \exp \left [ -\sum_{i}  ( \Delta A_\mu(x_i) )^2
\left ( -\frac{i}{2\hbar} \lambda_i \right ) \right ]
\end{eqnarray}
where $\lambda_i$ are the eigenvalues of the matrix $\left [\delta^2 \L/(\delta A_\mu(x_i) \delta A_\nu(y_j) ) \right ]$, and we have chosen the eigenvectors as a basis for performing the integral over the field configurations, $\int \prod \left ( \d A_\mu(x_i) \right )$.  Now let us perform the integration over each of the $\d A_\mu(x_i)$ to yield
\begin{eqnarray}
E_{A_\mu} &=& i \hbar \ln \prod_i \left ( -\frac{i}{2\hbar} \lambda_i \right )^{-1/2} = -\frac{i \hbar}{2} \ln  \prod_i \left (  \lambda_i \right ) - \frac{i\hbar}{2} \prod_i \left ( -\frac{i}{2\hbar} \right ) \\
	 &=& -\frac{i\hbar}{2} \ln \Det \left [
\frac{\delta^2 \L}{\delta A_\mu(x) \delta A_\nu(y) } \right ]  +
\mathrm{Constant~Terms}.
\end{eqnarray}
The constant terms absorb the constant prefactor in front of the integral, $-i/(2\hbar)$, as well as some divergent terms. The symbol $\Det$ denotes the functional determinant over both the spacetime and the spin space in the case of the Dirac fields. One subtlety is the plus sign in front of the functional derivative involving the Grassman fields $\bpsi(x)$ and $\psi(y)$.  Simply, the integral of 
\begin{equation}
\int d \bpsi d \psi \exp ( - \bpsi a \psi ) = \int d \bpsi d \psi  (1- \bpsi a \psi ) =\int d \bpsi d \psi  (1 + a\bpsi  \psi ) =  \int d \bpsi ( a\bpsi ) =  a ~\mathrm{not}~ \frac{2\pi}{a},
\label{eq:12}
\end{equation}
where the unexpected result comes from the anti-commuting nature of the fields. Performing the Legendre transformation yields an expression for the effective action to lowest order (one loop),
\begin{eqnarray}
\Gamma[ A^0_\mu ] =
\int \4dx \left ( - \frac{1}{4} F^0_{\mu\nu} F^{0,\mu\nu} \right ) 
+\frac{i\hbar}{2} \ln \Det \left [
\frac{\delta^2 \L}{\delta A_\mu(x) \delta A_\nu(y) } \right ] 
-i\hbar \ln \Det \left [ 
\frac{\delta^2 \L}{\delta \bpsi(x) \delta \psi(y) } \right ] 
\label{eq:13}
\end{eqnarray}
where we have dropped the constant terms from the expression. We have not been concerned with renormalizing the effective action so far, but the functional determinants are probably divergent.  Let us insist that the effective action vanishes as the classical field vanishes, so we have to subtract two terms corresponding to the functional determinants in the absence of an external field. This renormalizes the zero-point energy and yields,
\begin{eqnarray}
\Gamma[ A^0_\mu ] &=&
\int \4dx \left ( - \frac{1}{4} F^0_{\mu\nu} F^{0,\mu\nu} \right ) 
-\left . i\hbar \ln \Det \left [ \frac{\delta^2 \L}{\delta \bpsi(x) \delta \psi(y) } \right ] \right |_{A_\mu=A^0_\mu} 
+\left . i\hbar \ln \Det \left [ \frac{\delta^2 \L}{\delta \bpsi(x) \delta \psi(y) } \right ] \right |_{A_\mu=0} \\
&=& 
\int \4dx \left ( - \frac{1}{4} F^0_{\mu\nu} F^{0,\mu\nu} \right ) 
-i\hbar \ln \Det  \left [ \frac{ \Pi \!\!\!\! / - m }{ p \!\!\! / - m} \right ] 
\label{eq:14}
\end{eqnarray}
where
\begin{equation}
\Pi \!\!\!\! / = \gamma^\mu \Pi_\mu = 
i \gamma^\mu \pp{}{x^\mu} +e \gamma^\mu A^0_\mu  = 
\gamma^\mu p_\mu + e \gamma^\mu A^0_\mu .
\label{eq:15}
\end{equation}
The functional derivative of the Lagrangian with respect to the vector potential is same for all values of the classical vector potential as along as the fermionic classical field vanishes.  The effective action contains the classical Maxwell action of electrodynamics and an additional term that quantifies the effects of the vacuum fluctuations of the Dirac (here electron-positron) fields. 

We can use the linear algebra result, $\ln \Det A=\Tr \ln A$, to simplify the expression for the effective action further. We use the convention that $\Det$ and $\Tr$ span both coordinate and spin space, while $\tr$ and $\det$ just cover the spinorial components. (the analysis is this subsection builds upon \S 4-3-3 and \S 4-3-4 of \cite{Itzy80}), 
\begin{equation}
\Gamma[ A^0_\mu ] = \int \4dx \L_\mathrm{eff} =
\int \4dx \left ( - \frac{1}{4} F^0_{\mu\nu} F^{0,\mu\nu} \right ) 
-i\hbar
\Tr \ln  \left [ 
\frac{ \Pi \!\!\!\! / - m }{ p \!\!\! / - m}
 \right ]  
\label{eq:16}
\end{equation}
We would like to put the logarithm in a more manageable form.
\begin{equation}
\Tr \ln  \left [ 
\frac{ \Pi \!\!\!\! / - m }{ p \!\!\! / - m}
 \right ] =
\Tr \ln  \left [ 
\frac{ \Pi \!\!\!\! / + m }{ p \!\!\! / + m}
 \right ] = \frac{1}{2} \Tr \ln  \left [ 
\frac{ \Pi \!\!\!\!/^2 - m^2 }{ p \!\!\!/^2  - m^2}
 \right ] .
\label{eq:17}
\end{equation}
The first equality holds since the charge conjugation matrix $C$ satisfies $C\gamma_\mu C^{-1}=-\gamma_\mu^T$, so $C\Pi\!\!\!\!/C^{-1}=-\Pi\!\!\!\!/^T$ (similarly for $p\!\!\!/$), and the trace of an operator is invariant under transposition.  The second equality results from summing the first two expressions.  

\subsubsection{Effective Action : Proper-time Integration}
We use the identity
\begin{equation}
\ln \frac{a}{b} = \int_0^\infty \frac{\d s}{s} \left ( \exp is(b+i\epsilon)
- \exp is(a+i\epsilon) \right )
\label{eq:18}
\end{equation}
to expand the logarithm
\begin{equation}
\Tr \ln  \left [ 
\frac{ \Pi \!\!\!\! / - m }{ p \!\!\! / - m}
 \right ] = -\frac{1}{2} \int \4dx \int_0^\infty \frac{\d s}{s} e^{-i s m^2} e^{-\epsilon s}
\tr \biggr ( \left < x | \exp(i s \Pi \!\!\!\!/^2) | x \right >
- 
\left < x | \exp(i s p \!\!\!/^2) | x \right > \biggr )
\label{eq:19}
\end{equation}
and obtain the proper-time expression for the effective Lagrangian density \cite{Schw51},
\begin{equation}
\L_\mathrm{eff} = - \frac{1}{4} F^0_{\mu\nu} F^{0,\mu\nu}  
+\frac{i\hbar}{2} \int_0^\infty \frac{\d s}{s} e^{-i s m^2} e^{-\epsilon s} 
\tr \left ( \left < x | U(s) | x \right >
- \left < x | U_0(s) | x \right > \right )
\label{eq:20}
\end{equation}
where $U(s)$ is the time-evolution operator governed by the Hamiltonian,
\begin{equation}
{\cal H} = - \Pi\!\!\!\!/^2 = \Pi^\mu \Pi_\mu - \frac{1}{2} e \sigma^{\mu\nu} F^0_{\mu\nu}.
\label{eq:21}
\end{equation}
where $\sigma^{\mu\nu} = \frac{i}{2} [ \gamma^\mu, \gamma^\nu ]$.  $U_0(s)$ is the analogous operator for vanishing external fields. Eq.~\ref{eq:20} forms the basis of the worldline numerics techinque that facilitates the calculation of the effective action for arbitrary field configurations \cite{2002IJMPA..17..966G,2014arXiv1407.7490M}

\subsubsection{Results for a uniform field}
\label{sec:unieval}

We will select a particular frame and gauge to calculate the trace and obtain an expression for the effective Lagrangian from a uniform electromagnetic field.  We begin with
\begin{equation}
\tr \left ( \left < x | \exp(i s \Pi \!\!\!\!/^2) | x \right > \right ) =
\tr \left ( \left < x | \exp(i s \Pi^\mu \Pi_\mu) | x \right > \right ) \times 
\tr \left ( \left < x \left | \exp \left ( 
\frac{i}{2} e s \sigma^{\mu\nu} F^0_{\mu\nu} 
   \right ) \right | x \right > \right )
\label{eq:22}
\end{equation}
since $\sigma^{\mu\nu} F^0_{\mu\nu}$ commutes with $\Pi^\mu \Pi_\mu$ for constant fields.

We will choose a frame such that ${\bf E} \| {\bf B}$, and $a\equiv |{\bf E}|$ and $b \equiv |{\bf B}|$ where ${\bf E}$ and ${\bf B}$ are the classical electric and magnetic fields respectively.  In this frame, the eigenvalues of $\frac{i}{2} e s \sigma^{\mu\nu} F^0_{\mu\nu}$ are $\pm e s ( a \pm i b )$.  In a general frame, we can define 
\begin{equation}
(a + i b)^2 = ({\bf E} + i {\bf B} )^2 = |{\bf E}|^2 - |{\bf B}|^2 +
2 i {\bf E} \cdot {\bf B} 
\label{eq:23}
\end{equation}
which demonstrates that $a$ and $b$ are Lorentz invariants of the field. We obtain
\begin{equation}
\tr \left ( \left < x \left | \exp \left ( 
\frac{i}{2} e s \sigma^{\mu\nu} F^0_{\mu\nu} 
   \right ) \right | x \right > \right ) = 4 \cosh (e a s) \cos (e b s)
\label{eq:24}
\end{equation}
Evaluation of the first trace is more complicated.  With no loss of generality we can assume that the magnetic and electric fields point in the $z$-direction and select a gauge with $A^3 = - a t$ and $A^1 = -b y$; this yields,
\begin{equation}
\Pi^\mu \Pi_\mu = (P^0)^2 - (P^2)^2 - ( P^1 + e b X^2 )^2 
	- ( P^3 + e a X^0 )^2 
\label{eq:25}
\end{equation}  
Using the commutation relation $[x, p_x] = -i$, we can define the shift operator
\begin{equation}
e^{-i p_x c} f(x) e^{i p_x c} = f(x + c)
\label{eq:26}
\end{equation}
which allows us to write
\begin{equation}
\Pi^\mu \Pi_\mu = 
\exp \left ( -i \frac{P^2 P^1}{e b} - i \frac{P^0 P^3}{e a} \right ) 
\left [ (P^0)^2 - (P^2)^2 - ( e b X^2 )^2 
	- ( e a X^0 )^2  \right ]
\exp \left ( i \frac{P^2 P^1}{e b} + i \frac{P^0 P^3}{e a} \right ) 
\label{eq:27}
\end{equation}
To evalulate the trace itself we use the momemtum representation
\begin{eqnarray}
\tr \left ( \left < x | \exp(i s \Pi^\mu \Pi_\mu) | x \right > \right
) 
 &=& \int \frac{ \d p_3 \d p_1 }{(2 \pi)^4} \d p^0 \d p_0' \d p_2 \d p_2'
\exp \left [ i (p_0'-p_0) \left ( t + \frac{p_3}{ea} \right ) \right ]
\exp \left [ i (p_2'-p_2) \left ( y + \frac{p_2}{eb} \right ) \right ] 
\times \nonumber \\
& &~~~
\left < p_0 | \exp [i s ( P_0^2 - e^2 a^2 (X^0)^2 ) ] | p_0' \right >
\left < p_2 | \exp [-i s ( P_2^2 + e^2 b^2 (X^2)^2 ) ] | p_2' \right > \\
&=& \frac{e^2 a b}{(2 \pi)^2} 
\int_{-\infty}^\infty \d p_0
\left < p_0 | \exp [i s ( P_0^2 - e^2 a^2 (X^0)^2 ) ] | p_0' \right >  
\times \nonumber \\
& & ~~~
\int_{-\infty}^\infty \d p_2
\left < p_2 | \exp [-i s ( P_2^2 + e^2 b^2 (X^2)^2 ) ] | p_2 \right > 
\label{eq:28}
\end{eqnarray}
Let us examine the last of the two integrals in detail.
\begin{equation}
\int_{-\infty}^\infty \d p_2
\left < p_2 | \exp [-i s ( P_2^2 + e^2 b^2 (X^2)^2 ) ]|  p_2 \right >  
= \Tr \exp [-i s ( p^2 + e^2 b^2 x^2 ) ] = \Tr \exp [-2 i s {\cal H}]
\label{eq:29}
\end{equation}
where ${\cal H}$ is the Hamiltonian of a harmonic oscillator with unit mass and spring constant $k=e^2 b^2$.   Using the known eigenvalues of the system yields an expression for the integral,
\begin{equation}
\Tr \exp [-2 i {\cal H} s] = 
\sum_{n=0}^\infty \exp \left [-2 i e b s \left ( n +\frac{1}{2} \right
) \right ] = \frac{1}{2 i \sin (e b s) }.
\label{eq:30}
\end{equation}
The result for the first integral is similar except here $k=-e^2 a^2$, so the complete expression for 
\begin{equation}
\tr \left ( \left < x | \exp(i s \Pi \!\!\!\!/^2) | x \right > \right
) =
-i \frac{e^2 a b}{(2 \pi)^2} \coth (e a s ) \cot (e b s).
\label{eq:31}
 \end{equation}
Taking the limit of this expression as $a$ and $b$ vanish yields
\begin{equation}
\left < x | \exp(i s p \!\!\!/^2) | x \right > = 
-\frac{i}{(2\pi)^2} \frac{1}{s^2}.
\label{eq:32}
\end{equation}

\subsubsection{Effective Lagrangian in a constant field}

Substituting this result into \eqref{20} yields an expression for the effective Lagrangian,
\begin{equation}
\L_\mathrm{eff} = - \frac{1}{4} F^0_{\mu\nu} F^{0,\mu\nu} 
+\frac{\hbar}{2 (2 \pi)^2} \int_0^\infty \frac{\d s}{s} e^{-i s m^2} e^{-\epsilon s}
\left [
e^2 a b \coth (e a s) \cot (e b s) - \frac{1}{s^2} 
\right ]
\label{eq:33}
\end{equation}
For small values of $s$, the integrand diverges as $e^2 (a^2-b^2)/(3s)$.  Since this is proportional to the classical Lagrangian, we can absorb this infinite term through a renormalization which yields the renormalized expression for the effective Lagrangian,
\begin{equation}
\L_\mathrm{eff} = - \frac{1}{4} F^0_{\mu\nu} F^{0,\mu\nu}  
+\frac{\hbar}{8\pi^2} \int_0^\infty \frac{\d s}{s} e^{-i s m^2} e^{-\epsilon s} 
\left [
e^2 a b \coth (e a s) \cot (e b s) - \frac{1}{s^2} - \frac{1}{3} e^2
(a^2-b^2) 
\right ]
\label{eq:34}
\end{equation}
We can get a more convenient expression by performing a Wick rotation, and substituting $\zeta=s m^2$
\begin{equation}
\L_\mathrm{eff} = \frac{a^2-b^2}{2} + \frac{\alpha}{8\pi^2} B_\mathrm{\scriptsize QED}^2 \int_0^\infty 
\frac{d\zeta}{\zeta}
e^{-\zeta} 
\left [  \frac{a b}{B_\mathrm{\scriptsize QED}^2} \cot\left ( \zeta \frac{a}{B_\mathrm{\scriptsize QED}} \right) 
\coth \left (\zeta \frac{b}{B_\mathrm{\scriptsize QED}} \right) +
\frac{1}{\zeta^2} - \frac{1}{3} \frac{a^2-b^2}{B_\mathrm{\scriptsize QED}^2} \right ]
\label{eq:35}
\end{equation}
where $\alpha=e^2/(\hbar c)$ and we can now take $\epsilon\rightarrow 0$.

\subsection{Index of refraction for Low-Energy Photons}
\label{sec:index}

The index of refraction for low-energy photons is obtained most simply by defining the macroscopic fields as the generalized momenta conjugate to the fields \cite{Bere82},
\begin{equation}
{\bf D} = \pp{\lag}{\bf E} = {\bf E} + {\bf P}, 
{\bf H} = -\pp{\lag}{\bf B} = {\bf B} - {\bf M}, 
\label{eq:47}
\end{equation}
and linearizing these relations about the background field \cite{Adle71}. For an external magnetic field this yields \cite{Heyl97index}
\begin{eqnarray}
n_\perp &=& 1 - \frac{\alpha}{4\pi} X_1 \left(\frac{1}{\xi}\right)
\sin^2 \theta + {\cal O}\left[ \left (\frac{\alpha}{2\pi} \right )^2
\right] 
\label{eq:57}
\\
n_\| &=& 1 + \frac{\alpha}{4 \pi}
\left [ X_0^{(2)} \left ( \frac{1}{\xi} \right ) \xi^{-2} -
  X_0^{(1)} \left ( \frac{1}{\xi} \right ) \xi^{-1} \right ] \sin^2 \theta
+ {\cal O}\left[ \left (\frac{\alpha}{2\pi} \right )^2 \right ].
\label{eq:51}
\end{eqnarray}
where $\xi=B/B_\mathrm{\scriptsize QED}$, $n_\perp$ and $n_\|$ are the index if refraction for the perpendicular and parallel modes respectively (see below),
\begin{eqnarray}
X_1 \left ( \frac{1}{\xi} \right ) &=&
\frac{2}{3} \xi - \frac{1}{3} + 8 \left [ \ln A - \int_1^{1/(2\xi)+1} \ln \Gamma(v) \d v \right ] + \frac{2}{3}  \Psi \left ( \frac{1}{2\xi} \right ) 
\nonumber \\*
& & ~~~
+ \frac{1}{\xi} \left [ 
2\ln \Gamma \left ( \frac{1}{2\xi} \right ) - 
3 \ln \xi + \ln \left ( \frac{\pi}{4} \right ) - 2  \right ] - \frac{1}{2\xi^2} \; ,\\
 X_0^{(2)} \left ( \frac{1}{\xi} \right ) \xi^{-2} -
  X_0^{(1)} \left ( \frac{1}{\xi} \right ) \xi^{-1} &=& \frac{2}{3} + 
\frac{1}{\xi} \left [ -2 \ln \Gamma \left ( \frac{1}{2\xi} \right ) + \ln \xi + \ln 4\pi + 1  \right ] + \nonumber \\* 
& & ~~~ \frac{1}{\xi^2} \left [ \Psi \left ( \frac{1}{2\xi} \right ) - 1 \right ] 
\label{eq:49}
\end{eqnarray}
and $\ln A = \frac{1}{12} - \zeta^{(1)}(-1) \approx 0.248754477$.  The functions $X_0(1/\xi)$ and $X_1(1/\xi)$ are related to the effective action and its derivative with respect to $a$ in the limit of $a \rightarrow 0$ \cite{Heyl97hesplit}.

Our naming convention is the following: if $\epsilon^{\mu\nu\alpha\beta} F_{\mu\nu} F_{\alpha\beta} = 0$ or $\efi \cdot \bfi =0$, the photon is in the parallel mode, otherwise it is in the perpendicular mode, where $F_{\mu\nu}$ is the sum of field tensors of the wave and external field.  In the weak field limit, $n-1 \propto \xi^2$; while in the strong field limit $n_\perp-1 \propto \xi$ and $n_\|$ approaches a constant \cite{Heyl97index}.  In particular,  in the weak field we have
\begin{equation}
n_\perp = 1 + \frac{\alpha}{4\pi} \frac{14}{45} \xi^2 \sin^2\theta ~~  n_\| = 1 + \frac{\alpha}{4\pi} \frac{8}{45} \xi^2 \sin^2 \theta
\end{equation}
and a birefringence of 
\begin{equation}
n_\perp - n_\| = \frac{\alpha}{4\pi} \frac{2}{15} \xi^2 \sin^2 \theta.
\end{equation}

\subsection{Propagation through a Birefringent Medium}
\label{sec:propagation}

As polarized radiation propagates through a birefringent medium, the direction of polarization changes.  If we use the Stokes parameters $S_0, S_1, S_2$ and $S_3$ ($I, Q, U, V$), we can define the normalized Stokes vector $\vec s=(S_1,S_2,S_3)/S_0$ and follow the evolution of the polarization using 
\cite{1983JOSA...73.1719K} \begin{equation}
\dd{\vec s}{\lambda} =  \hat \Omega \times \vec s~\mathrm{, where}~
|\hat \Omega|=|\Delta k| = \frac{\alpha}{15} \frac{\nu}{c} \left ( \frac{B_\perp}{B_\mathrm{\scriptsize QED}} \right )^2
\label{eq:54}
\end{equation}
and where $\lambda$ measures the length of the photon path and $\nu$ is the frequency of the radiation.  The value of $\Delta k$ is the difference in the wavenumber for the two polarization states, and the final equality holds in the weak-field limit of QED. The direction of $\hat \Omega$ points toward the polarization of the perpendicular mode on the Poincar\'e sphere of polarization states.  

This equation may seem more familiar if one considers the Faraday rotation of polarized light passing through a weakly magnetized plasma. In this case, $\hat \Omega$ points toward the $s_3$-direction, corresponding to the  circular polarization, so the polarization direction of linearly polarized light will rotate.  In general, if the direction of $\hat \Omega$ is constant, the vector $\vec s$ will circle the direction of $\hat \Omega$.
If the magnetic field is sufficiently strong so that $|\hat \Omega|$ is sufficiently large, the vector $\vec s$ will circle the direction of $\hat \Omega$ even in the case in which $\hat \Omega$ changes direction and magnitude, if it does so sufficiently gradually.  In particular, if the polarization state is initially parallel to $\hat \Omega$, that is, the initial polarization is parallel or perpendicular to the magnetic field, the polarization state will remain nearly parallel to $\hat \Omega $ as long as \cite{Heyl99polar}
\begin{equation}
\left | \hat \Omega \left ( \dd{\ln |\hat \Omega|}{\lambda} \right )^{-1} \right | \geq 0.5.
\label{eq:56}
\end{equation}
If this condition holds, the polarization states evolve adiabatically, and the polarization direction will follow the direction of the birefringence. 

\begin{figure}
\centering
\begin{tikzpicture}
\draw (0,0) node {\includegraphics[width=0.7\textwidth]{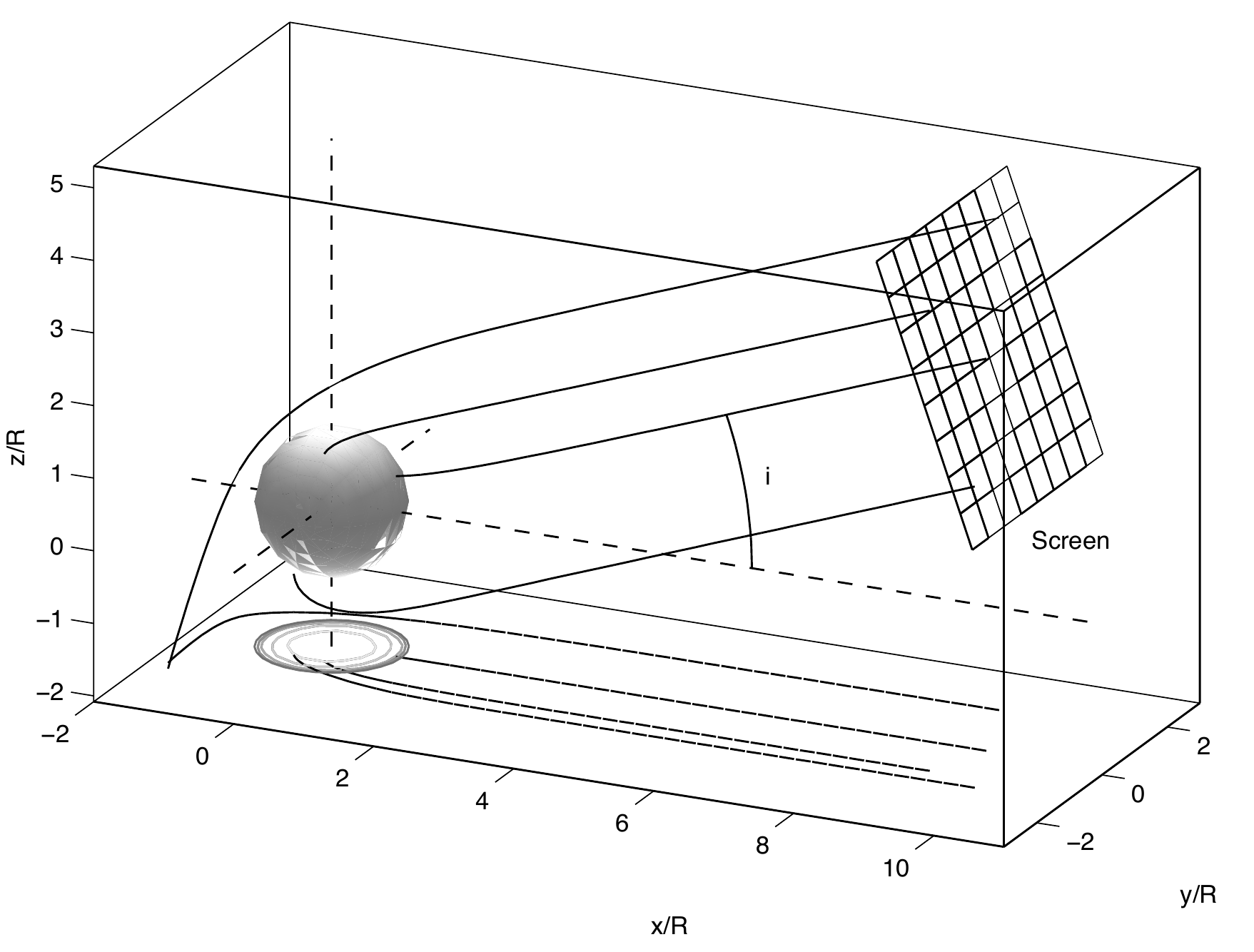}};
\filldraw[fill=white,draw=white] (3.25,-0.8) -- (3,-0.8) -- (2.1,1.6) -- (3.25,1.4) -- cycle;
\filldraw[fill=white,draw=white] (3.3,-0.8) -- (4.7,-0.8) -- (4.7,2.45) -- (3.3,1.4) -- cycle;
\filldraw[fill=white,draw=white] (4.7,2.5) -- (3.3,1.47) -- (1.9,1.7) -- (3.3,2.8) -- cycle;
\draw (-2,0.1) node {\includegraphics[width=0.05\textwidth,height=0.1\textwidth,angle=20]{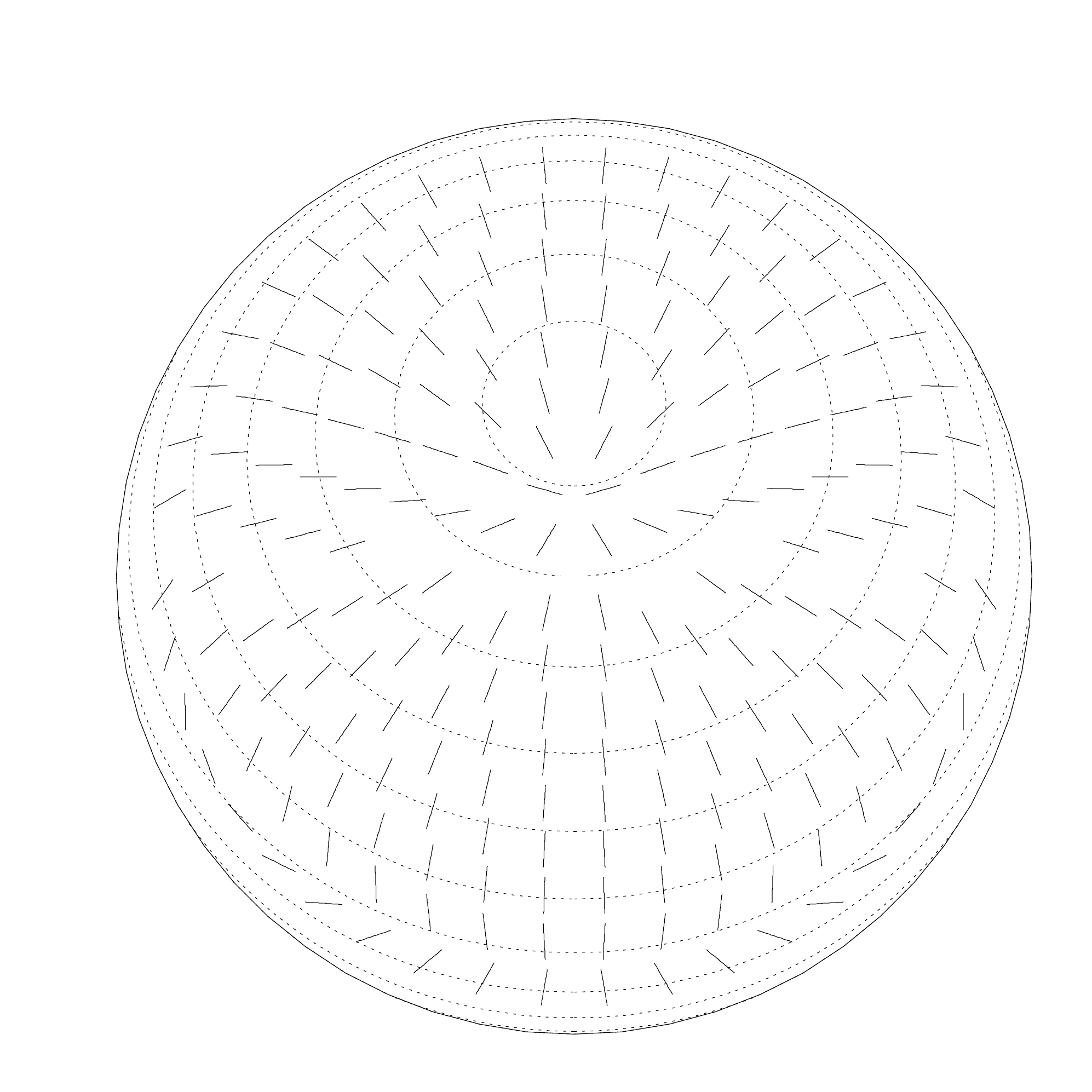}}
(-2,1) node {Small $r_\mathrm{PL}$}
(3.15,1.1) node {\includegraphics[width=0.09\textwidth,height=0.18\textwidth,angle=20]{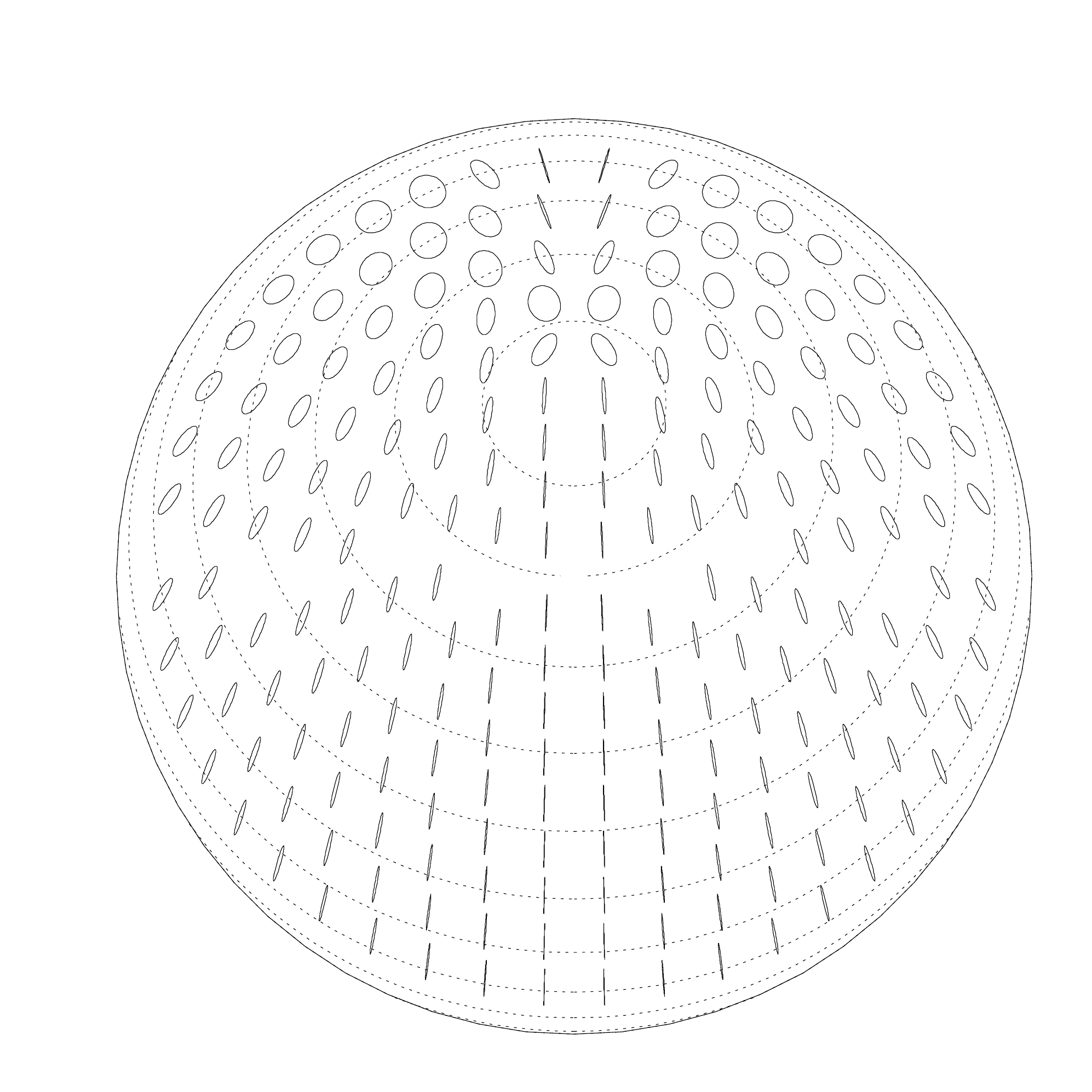}} 
(3.15,2.6) node {Large $r_\mathrm{PL}$};
\end{tikzpicture}
 \caption{Radiation leaving the surface of a neutron star follows geodesics so that the bundle of rays that reaches the distant observer is approximately cylindrical.  The three-dimensional coordinates ($x,y,z$) are given in terms of the radius of the neutron star ($R$). If the polarization-limiting radius is small, the final polarization will reflect the magnetic field structure near to the star where the bundle covers a large fraction of hemisphere so the field structure varies a lot over the bundle at this point, and the final polarization will also vary a lot over the image.  On the other hand, if the polarization-limiting radius is large, the field structure over the ray bundle is simpler, and the polarization direction will not vary much over the image.  Since the magnetic field is assumed to be that of a dipole  (aligned with the $z$-axis), it has axial symmetry and different images will be distinguishable by the observer's magnetic inclination angle $i$. Adapted from \cite{Shav98lens}}
\label{ray_tracing_fig}
\end{figure}

If we specialize to the dipole field surrounding a neutron star where $B \approx \mu r^{-3}$, where $\mu$ is the magnetic dipole moment of the star and $r$ is the distance from the center of the star, we find the following condition
\begin{equation}
\left | \frac{\alpha}{15} \frac{\nu}{c} \frac{\mu^2 \sin^2 \beta}{r^6 B^2_\mathrm{\scriptsize QED}} \frac{r}{6}  \right | \geq 0.5
\label{eq:56}
\end{equation}
where $\beta$ is the angle between the dipole axis and the line of sight. If we define the polarization-limiting radius ($r_\mathrm{\scriptsize PL}$) to be the distance at which the equality holds, we find that the polarization will follow the direction of magnetic field out to  
\begin{equation}
r_\mathrm{\scriptsize PL} = \left ( \frac{\alpha}{45}
 \frac{\nu}{c} \right )^{1/5} \left ( \frac{\mu}{B_\mathrm{\scriptsize QED}} \sin
 \beta \right )^{2/5} \approx  1.2 \times 10^{7} \left
 ( \frac{\mu}{10^{30}~\mathrm{G~cm}^3} \right )^{2/5} \left (
 \frac{\nu}{10^{17}~\mathrm{Hz}} \right)^{1/5} \left ( \sin \beta
 \right)^{2/5} \mathrm{cm.}
 \label{eq:52}
\end{equation}
Fig.~\ref{ray_tracing_fig} illustrates the propagation of radiation away from the surface of the neutron star toward a distant observer.  For X-ray photons coming from near the surface of a neutron star with a surface field of $10^{12}$~G, the polarization-limiting radius is much larger than the star, according to Eq.~\ref{eq:52}, so the observed polarization of the photons will reflect the direction of the magnetic field at a large distance from the star and not at the surface.  For a much more weakly magnetized star, the polarization-limiting radius will be comparable to the radius of the star, so the observed polarization will reflect the field structure close to the star.

\begin{figure}
\includegraphics[width=0.5\textwidth]{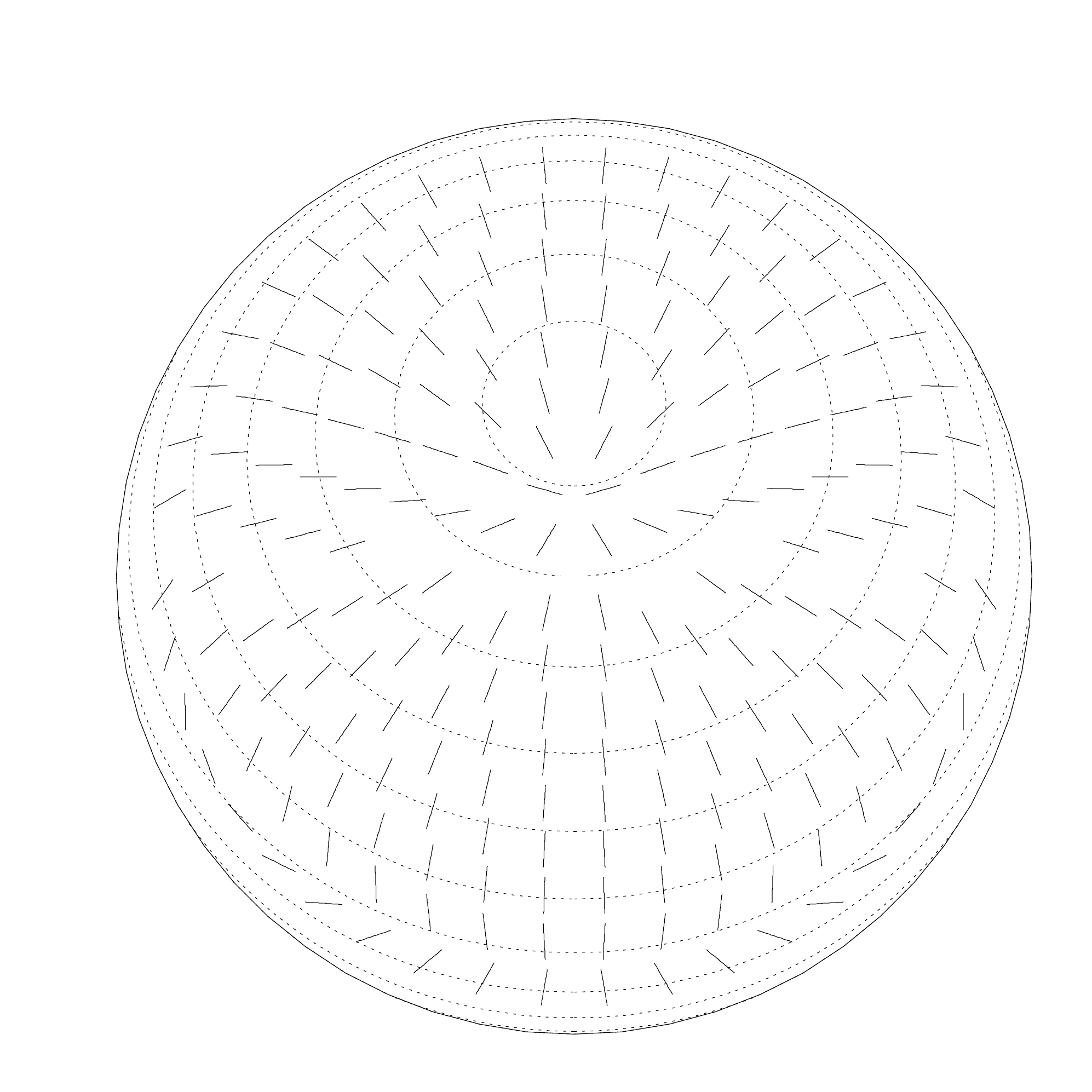}
\includegraphics[width=0.5\textwidth]{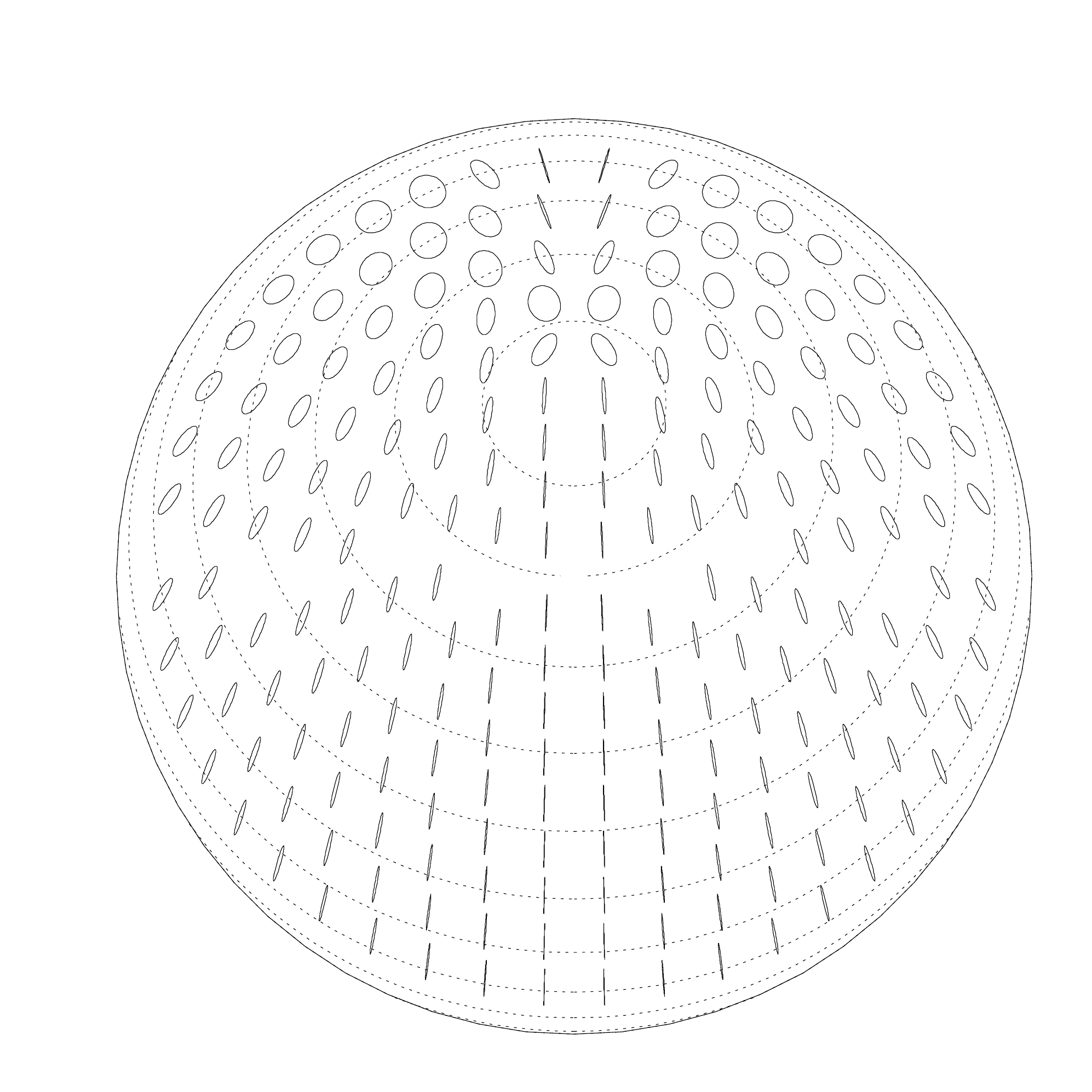}
\caption{ The polarized emission map of a neutron star overlaid on the apparent image of the NS.  The left panel depicts the observed map of polarization directions if one assumes that the surface emits only in the ordinary mode (parallel to the local field direction) and neglects the vacuum birefringence induced by QED.  The right panel shows the polarization map including birefringence for a frequency of $\nu =(\mu/(10^{30}\mathrm{G ~cm}^3))^{-2} 10^{17}$~Hz. The ellipses and short lines describe the polarization of a light ray originating from the surface element beneath them. The lines and the major axes of the ellipses point towards the direction of the linear component of the polarization  direction. The minor to major axis ratio provides the amount of circular polarization ($s_3$).  In both maps, the large dashed curves are lines of constant magnetic latitude (separated by $15^{\circ}$). The observer's line of sight makes an angle of $30^\circ$ with the dipole axis.  For comparison, if one assumes that the entire surface is emitting fully polarized radiation, the net linear polarization on the left 13\% while it is 70\% on the right. 
  }
  \label{fig:pol_map}
  \end{figure}

Fig.~\ref{fig:pol_map} depicts the final polarization states across the image of the neutron star surface assuming that the radiation is initially in the ordinary mode, that is, the electric field is parallel to the local magnetic field.  The left panel shows the case where the vacuum birefringence is neglected, and the right panel show the case where the surface field is about $10^{12}$~G and the frequency is $10^{17}$~Hz or an energy of about 0.4~keV. This is appropriate for a thermally emitting neutron star such as one of the X-ray dim neutron stars (XDINS). The effect of the vacuum polarization is to comb the polarization direction to be aligned with the direction of the magnetic axis of the star and dramatically increase the observed total polarization from about 13\% to about 70\%.  For more strongly magnetized neutron stars the effect is more dramatic. 

The studies of the effects of QED on the net polarization from astrophysical sources are more mature for the magnetars and the XDINS. However, it may be important for other sources as well.  Tab.~\ref{tab:sources} lists several classes of possible sources.  For the magnetars and XDINS, the polarization-limiting radius is much larger than the radius of the star. Because we expect the emission in these sources to come from a larger region of the stellar surface or magnetosphere (in the case of the non-thermal emission from magnetars \cite{2014MNRAS.438.1686T}), we expect a large increase in the observed polarization fraction due to QED. Although the ratio of the polarization-limiting radius to the stellar radius is also large for the X-ray pulsars (XRP), as we shall see in \S~\ref{sec:results}, the effect for these objects is more subtle. The QED effects for more weakly magnetized stars such as millisecond XRPs (ms XRPs) and strongly magnetized white dwarfs have not yet been explored. However, QED has been found to be generically important in the interpretation of X-ray polarization from accreting black holes \cite[][and Caiazzo \& Heyl this volume]{Caia18bh}.
 \begin{table}
 \caption{The expected polarization-limiting radii for various sources; the typical observing times are for eXTP at 2-8~keV for the magnetars (4U~0142+61) and XRPs (Her~X-1) from the text and for RedSOX \cite{SPIE_REDSoX} at 0.2-0.8~keV for RX J1856.5-3754 to make a four-sigma detection.}
 \label{tab:sources}
{ \centering
    \begin{tabular}{l|cccccc}
    \hline
    & Radius [cm]      & Magnetic Field [G]    & $\mu$ [G~cm$^{-3}$]& $r_\mathrm{pl}$ at 4~keV [cm] & $r_\mathrm{pl}/R$  & $t_\mathrm{obs}$ 
  \\
    \hline
    Magnetar       & $10^6$ & $10^{15}$ & $10^{33}$ & $3.0 \times 10^8$ & 300 & 10~ks  \\
    XDINS          & $10^6$ & $10^{13}$ & $10^{31}$ & $4.7 \times 10^7$ & 50$^*$ & 1~ks   \\
    XRP            & $10^6$ & $10^{12}$ & $10^{30}$ & $1.9 \times 10^7$ & 20  & 100~ks  \\
    ms XRP         & $10^6$ & $10^{9}~$  & $10^{27}$ & $1.2 \times 10^6$ & 1.2  & \\
    AM Her         & $10^9$ & $10^{8}~$  & $10^{35}$ & $1.9 \times 10^9$ & 1.9 & \\
    Black Hole     & $10^{6+}$ &    ?    &   N/A     &   \multicolumn{2}{c}{See \cite{Caia18bh}} & \\ \\
  \end{tabular}
  }
  $^*$XDINS have little emission at 4~keV, and therefore will be difficult to observe with eXTP and IXPE.  The value of $r_\mathrm{pl}/R$ at 0.4~keV is 32, so vacuum birefringence is important for observations with soft-X-ray polarimeters.
\end{table}
\section{Results}
\label{sec:results}

\begin{figure}
\centering
\includegraphics[width=0.7\textwidth]{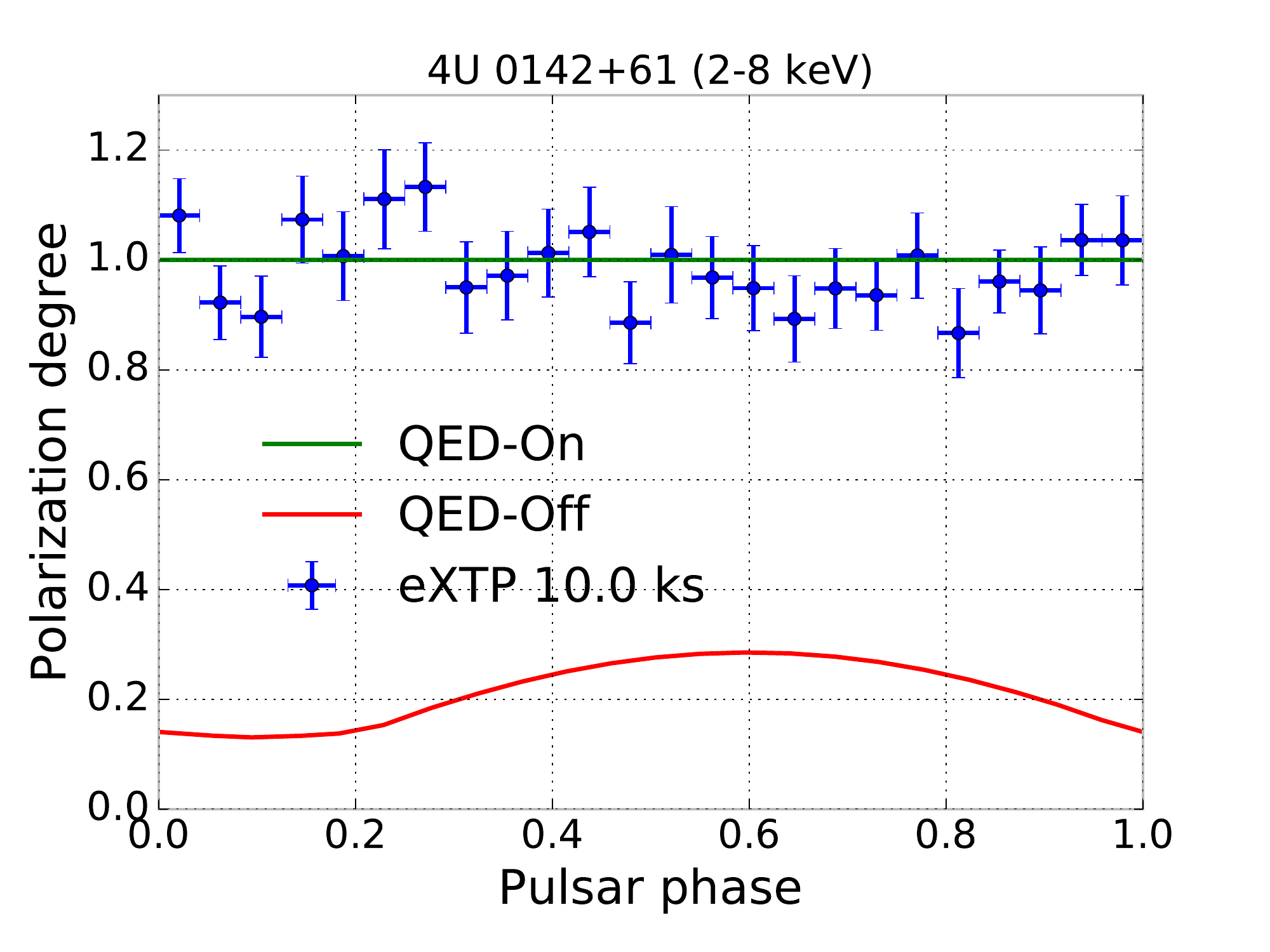}
\caption{The expected degree of polarization from the magnetar 4U~0142+61 from the thermal emission, assuming a two-black-body model in which the radiation is initially fully polarized in the perpendicular mode.  The crosses give the results for a short simulated observation with eXTP (10~ks). The results for IXPE are similar with a 30-ks observation.}
\label{fig:magnetar}
\end{figure}

To demonstrate the importance of QED vacuum birefringence on the X-ray polarization from neutron stars, we focus on two particular objects: the magnetar 4U~0142+61 and the X-ray pulsar Hercules X-1.  For the magnetar, the effect is especially dramatic.  The magnetar 4U~0142+61 exhibits a X-ray spectrum from 2-8~keV (the range of sensitivity of IXPE and eXTP) that is dominated by thermal emission \cite{2015ApJ...808...32T}.  As argued by \cite[e.g.][]{Heyl01polar,2003ApJ...599.1293H,2005astro.ph..2351H,2009MNRAS.399.1523V,2016arXiv160501281P}, this thermal emission is expected to be nearly fully polarized as it is emitted.  We have used the phase-resolved spectral fits of \cite{2015ApJ...808...32T} to simulate the polarized spectra in XIMPOL \cite{2016cosp...41E.129B} for a 10.0-ks observation with eXTP.  Because we expect, when QED is included, that the final polarization will follow the magnetic field direction at a large distance from the star, the final polarization fraction with QED will be near 100\% \cite{Heyl01polar}.  On the other hand without QED, the expected final polarization will be lower as apparent from the left panel of Fig.~\ref{fig:pol_map} and will depend on where the emission originates.  Using the fits of \cite{2015ApJ...808...32T} as a guide,  we assume that the hot blackbody component originates from a hot-spot of ten degree radius and that the cool blackbody originates from the entire stellar surface.  We assume that the hard-power-law component will not contribute in the 2-8~keV range. Even without QED birefringence, we expect the first component to remain highly polarized because it originates from a small region near the polar cap where the projected field is well aligned (see the left panel of Fig.~\ref{fig:pol_map}), but the second component will not, because it comes from the rest of the surface.  Fig.~\ref{fig:magnetar} presents the results of this calculation along with a short simulated observation with eXTP.  Even a relatively short observation of this magnetar can detect the effect of QED vacuum birefringence dramatically.  The emission from other magnetars such as AXP~1RXS~J170849.0-400910 is dominated by non-thermal magnetospheric emission even in the 2-8~keV band of eXTP and IXPE so to understand the X-ray polarization from these objects requires a magnetospheric emission model \cite[e.g.][]{2014MNRAS.438.1686T}.  However, because the QED-induced polarization-limiting radius is much larger than the emission region even in this case, the results are similar: QED dramatically increases the expected polarization of these objects.
\begin{figure}
 \includegraphics[height=0.5\textwidth]{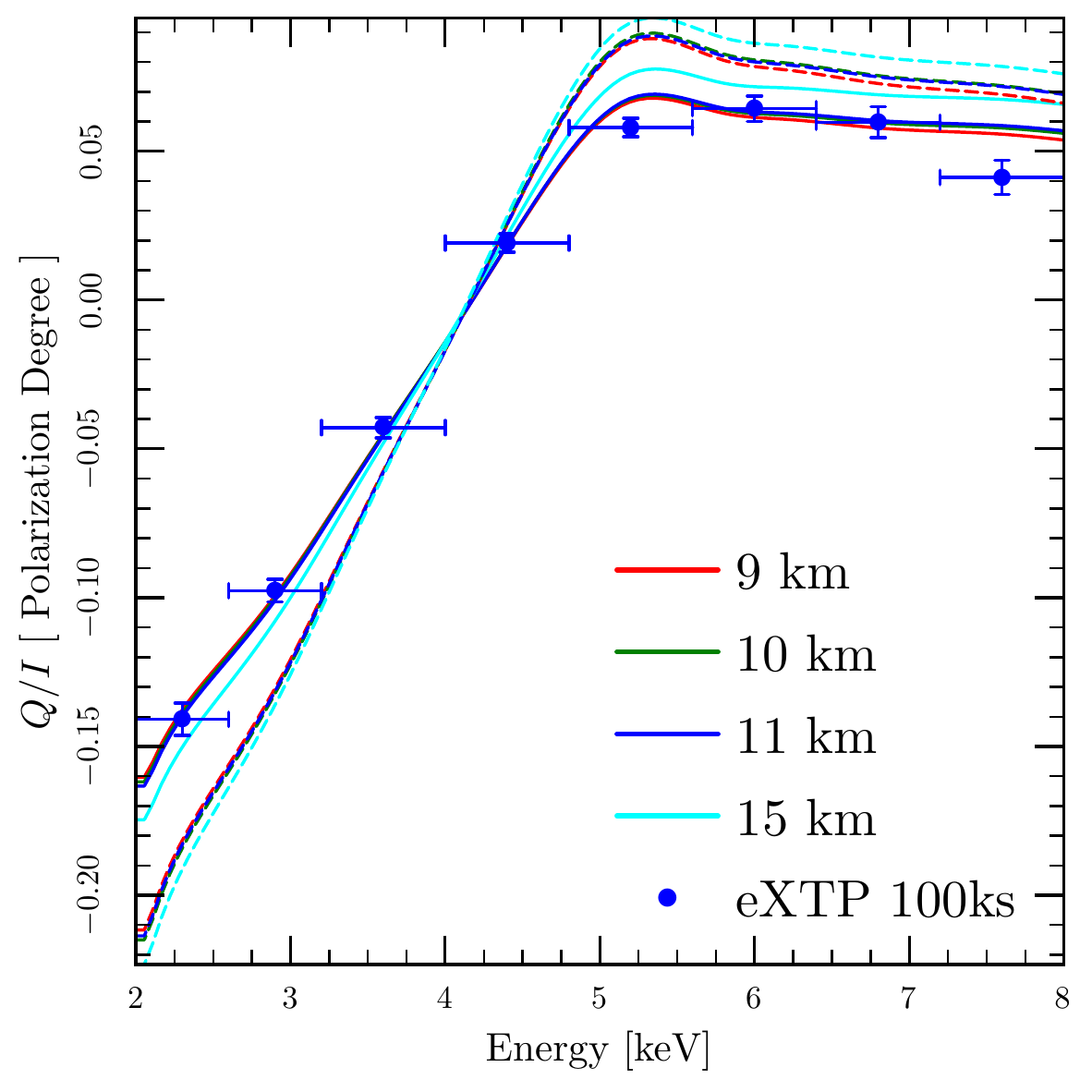}
    \begin{tikzpicture}
      \draw (0,0) node { \includegraphics[height=0.5\textwidth,,clip,trim=0.3in 0 0 0]{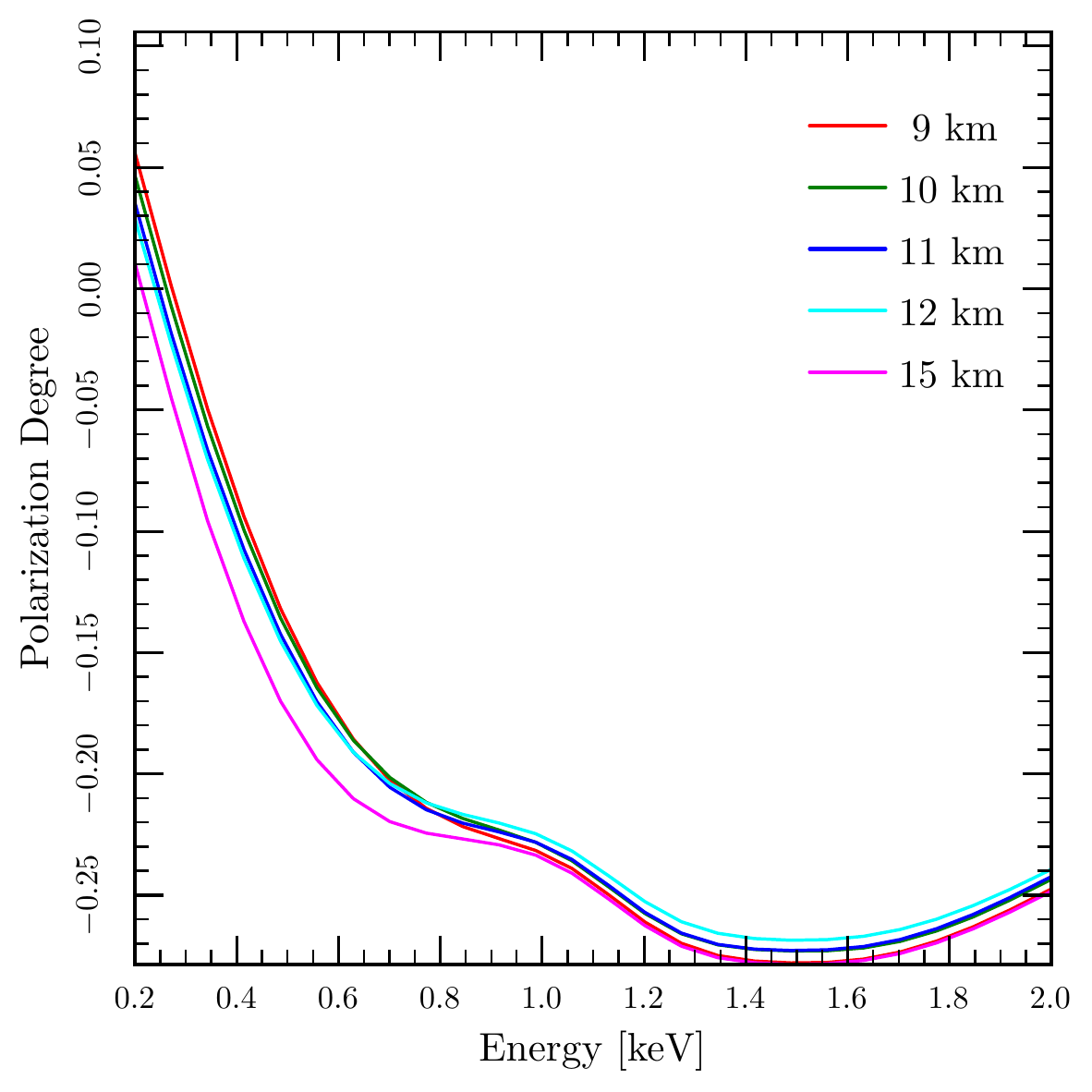} } (0.75,0.9) node { \includegraphics[width=0.33\textwidth]{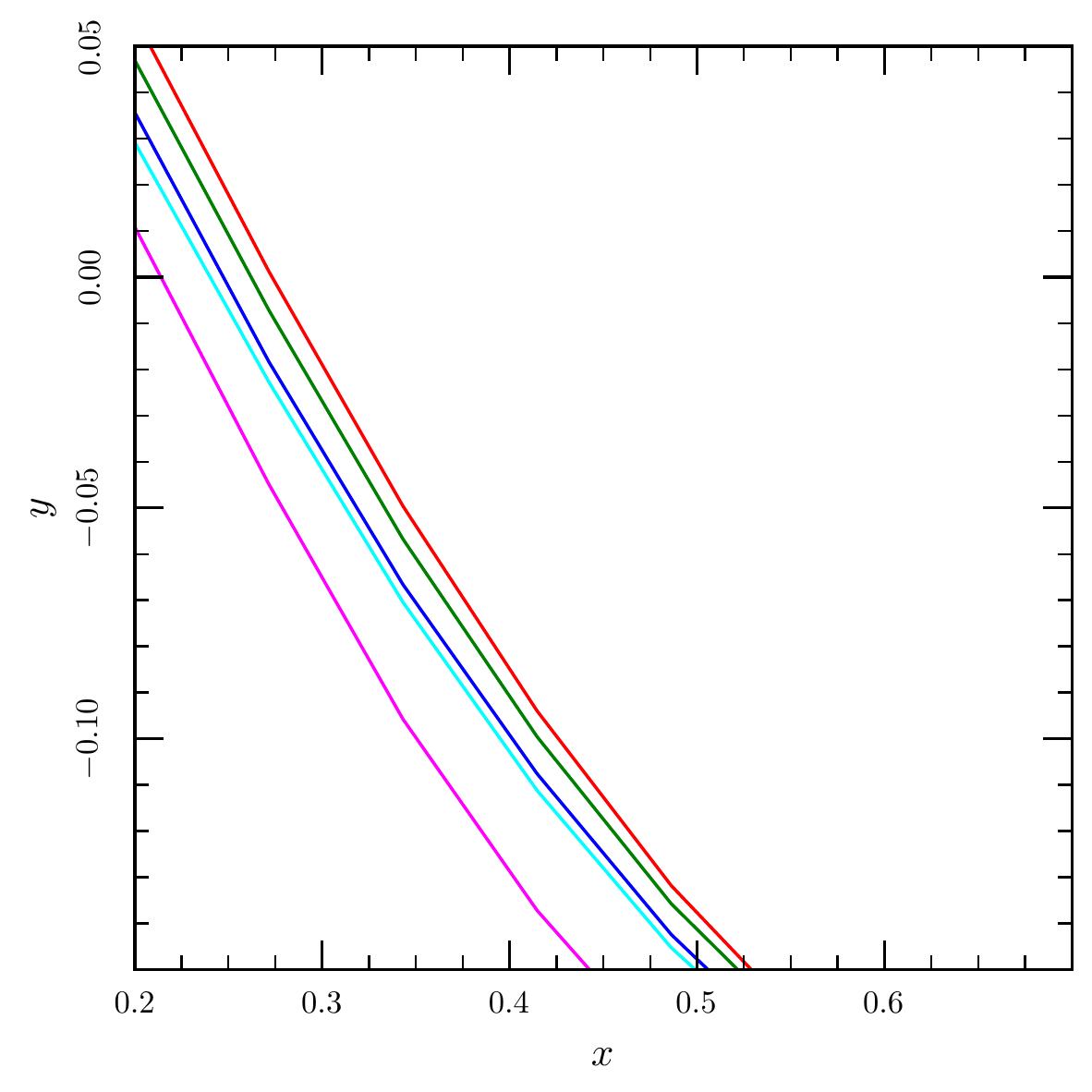} };
      \filldraw[fill=white,draw=white] (-2, 0.9) rectangle (-1.55,1.3);
      \filldraw[fill=white,draw=white] (0.9,-2) rectangle (1.3,-1.4);
      \end{tikzpicture}
       \caption{The polarization from Her~X-1 as function of photon energy using the emission models of \cite{1987PASJ...39..781K,1988ApJ...324.1056M}, averaged over the rotation of the pulsar.  A positive value of the polarization degree indicates that the polarization direction is perpendicular to the projection of the rotation axis onto the plane of the sky.  In the left panel, the dashed lines give the result without vacuum birefringence (we have reversed the polarization direction in this case for ease of comparison, see the text) and the solid lines include the QED effect.  A positive value of $Q/I$ indicates that the net polarization is perpendicular to the projected spin axis of the stars in the QED-on case, and parallel in the QED-off case.  The crosses give the results of a 100-ks simulated observation with eXTP. The results for IXPE are similar with a 300-ks observation.}
    \label{fig:hex1}
\end{figure}

The second object that we consider is the X-ray pulsar Her~X-1.  We consider that the radiation comes from a slab at the polar cap with a radius of six degrees. In reality the X-ray emission comes from both the accretion disk and from near the star, and the contribution from the disk is expected to be more important at lower energies.   Furthermore, even the structure of the emission near the star is uncertain: it could be a slab or an accretion column, and there are non-thermal contributions as well.  For simplicity here we just treat the slab geometry and argue that with phase-resolved polarimetry, the geometric ambiguities and the location of various emission regions can be disentangled. We use the results of \cite{1988ApJ...324.1056M} to estimate the total flux from the region and the results of \cite{1983JOSA...73.1719K} to estimate the polarized fraction and the intensity as a function of direction from the slab.  We choose a geometry that results in a large pulsed fraction. We again use XIMPOL \cite{2016cosp...41E.129B} to perform the instrumental simulation including effects of QED birefringence by integrating Eq.~\ref{eq:54} through the magnetosphere. We assume a range of possible stellar radii from 9 to 15~km.  From Fig.~\ref{fig:pol_map}, we see that when the emission is restricted to the region near the magnetic pole, the effect of QED is more subtle than for the magnetar.  The projected magnetic field direction is well aligned near the pole even without vacuum birefringence in the left panel.  The final polarization in the right panel is well aligned in general but substantial circular polarization can be generated near the pole \cite{2009MNRAS.398..515W}.  Because the instrument only detects linear polarization, the circularly polarized radiation does not contribute to the observed polarization fraction, so the net effect of the vacuum birefringence is to reduce the polarization fraction integrated over the emission region and the rotational phase as shown by the left panel of Fig.~\ref{fig:hex1}.  The solid lines trace the extent of linear polarization including QED, and the dashed lines neglect it.  Across the 2-8-keV energy range, the trend in $Q/I$ can be explained by looking at the opacities in the atmosphere.  At low energies the opacity for photons in the parallel polarization is larger than for the perpendicular polarization, so the bulk of the radiation emerges in the perpendicular mode.  However, as the photon energy approaches the cyclotron energy (here taken to be 38~keV), the opacity for the perpendicular mode increases and one gets more emission in the parallel mode.

The right panel of Fig.~\ref{fig:hex1} really highlights the role of QED.  Below about 400~eV the polarization swings as well.  All of the photons coming through the atmosphere pass through the vacuum resonance region in which the linear contribution to the birefringence from QED cancels that of the plasma \cite{2003MNRAS.338..233H}. In this region the value of $\hat \Omega$ swings from pointing along a particular direction in the $s_1-s_2-$plane up toward $s_3$ and back onto the $s_1-s_2-$plane in the opposite direction.  As we know from Eq.~\ref{eq:56} if this happens slowly enough the photon polarization will follow the direction of $\hat \Omega$, and this is in fact what happens for photons with energies greater than about 350~eV. The polarization state is switched from perpendicular to parallel.  Because this QED effect reverses the polarization direction for all of the photons above this energy, we have switched the sign of the value of $Q/I$ for QED-off case for ease of comparison.  Without an independent measurement of the projection of the spin axis of the star into the plane of the sky, measurements just in the 2-8-keV band cannot measure this polarization flip.  The key effect of the QED birefringence in this harder band is to slightly reduce the polarization fraction.

We can apply Eq.~\ref{eq:56} to determine how the conditions in the atmosphere, in particular the density scale height, determine the critical energy above which the polarization switches as the radiation passes through the vacuum resonance \cite{2003MNRAS.338..233H}.  In the case of Her~X-1 we estimate the temperature of the atmosphere to be 8~keV \cite{1988ApJ...324.1056M}.  With the temperature fixed, the density scale height only depends on the composition of the atmosphere which is that of the donor star and the surface gravity; therefore, the photon energy at which the polarization direction flips depends on the surface gravity and can be used to measure the radius of the star.  For the curves in Fig.~\ref{fig:hex1} we have assumed a mass of 1.4~M$_\odot$ for the neutron star.
%
\section{Discussion}

We have presented an {\em ab initio} derivation of the effective Lagrangian of quantum electrodynamics and the induced vacuum birefringence.  We have found that QED vacuum birefringence dramatically increases the total polarization fraction of the the thermal emission from a magnetar. The bright, hot magnetar 4U~0142+61 is an excellent candidate to probe the QED vacuum with the current generation of X-ray polarimetry missions: IXPE and eXTP. We presented the first calculation of the expected polarization from an X-ray pulsar, Her~X-1, to include vacuum birefringence in the magnetosphere.  For these more weakly magnetized objects, the effect of QED is more subtle.  In the IXPE and eXTP energy band, QED reduces the total polarization fraction slightly and the resulting extent of polarization has a modest dependence on the radius of the neutron star. If the emission at lower energies indeed comes from the stellar atmosphere, the direction of the polarization will swing by ninety degrees at about 400~keV.  This occurs because higher energy photons can pass through the vacuum resonance in the stellar atmosphere adiabatically \cite{2003MNRAS.338..233H}.  The transition energy between adiabatic and non-adiabatic evolution depends on the density scale height in the atmosphere, so a measurement of this energy could constrain the radius of the neutron star.

\section{Materials and Methods}

We integrate the equations of the polarization evolution through the neutron-star magnetosphere using an adaptive Runge-Kutta method as outlined in \cite{Heyl01qed}.  The software is available at \href{https://github.com/UBC-Astrophysics/QEDSurface}{https://github.com/UBC-Astrophysics/QEDSurface}.  We use the XIMPOL package \cite{2016cosp...41E.129B} to perform the instrumental simulations (\href{https://github.com/lucabaldini/ximpol}{https://github.com/lucabaldini/ximpol}).



\vspace{6pt} 


\acknowledgments{We used the NASA ADS service, arXiv.org and SIMBAD. This work was supported by a Discovery Grant from the Natural Sciences and Engineering Research Council of Canada, the Canadian Foundation for Innovation and the British Columbia Knowledge Development Fund. I.C.\ is supported by a Four-Year-Fellowship from the University of British Columbia.}

\authorcontributions{J.H.\ and I.C.\ conceived the calculations for 4U~0142+61 and Her~X-1.  J.H.\ performed the calculations for 4U~0142+61 and the derivations in S~\ref{sec:effaction_derivation}.   I.C.\ performed the calculations for Her~X-1.  J.H.\ wrote the bulk of the paper.}

\conflictofinterests{The authors declare no conflict of interest.} 





\bibliographystyle{mdpi}
\bibliography{mine,qed}

\begin{thebibliography}{-------}
\providecommand{\natexlab}[1]{#1}

\bibitem[Heisenberg and Euler(1936)]{Heis36}
Heisenberg, W.; Euler, H.
\newblock {\em Z. Physik} {\bf 1936}, {\em 98},~714.

\bibitem[Weisskopf(1936)]{Weis36}
Weisskopf, V.S.
\newblock Uber die elektrodynamik des vakuums auf grund der quantentheorie des
  elektrons.
\newblock {\em Kongelige Danske Videnskaberns Selskab, Mathematisk-Fysiske
  Meddelelser} {\bf 1936}, {\em 14},~1.

\bibitem[{Dirac}(1928)]{1928RSPSA.117..610D}
{Dirac}, P.A.M.
\newblock {The Quantum Theory of the Electron}.
\newblock {\em Proceedings of the Royal Society of London Series A} {\bf 1928},
  {\em 117},~610--624.

\bibitem[{Dirac}(1931)]{1931RSPSA.133...60D}
{Dirac}, P.A.M.
\newblock {Quantised Singularities in the Electromagnetic Field}.
\newblock {\em Proceedings of the Royal Society of London Series A} {\bf 1931},
  {\em 133},~60--72.

\bibitem[{Anderson}(1933)]{1933PhRv...43..491A}
{Anderson}, C.D.
\newblock {The Positive Electron}.
\newblock {\em Physical Review} {\bf 1933}, {\em 43},~491--494.

\bibitem[{Mignani} \em{et~al.}(2017){Mignani}, {Testa}, {Gonz{\'a}lez
  Caniulef}, {Taverna}, {Turolla}, {Zane}, and {Wu}]{2017MNRAS.465..492M}
{Mignani}, R.P.; {Testa}, V.; {Gonz{\'a}lez Caniulef}, D.; {Taverna}, R.;
  {Turolla}, R.; {Zane}, S.; {Wu}, K.
\newblock {Evidence for vacuum birefringence from the first optical-polarimetry
  measurement of the isolated neutron star RX J1856.5-3754}.
\newblock {\em \mnras} {\bf 2017}, {\em 465},~492--500,
  \href{http://xxx.lanl.gov/abs/1610.08323}{{\normalfont
  [arXiv:astro-ph.HE/1610.08323]}}.

\bibitem[Schwinger(1951)]{Schw51}
Schwinger, J.
\newblock On Gauge Invariance and Vacuum Polarization.
\newblock {\em Physical Review} {\bf 1951}, {\em 82},~664.

\bibitem[{Zavattini} \em{et~al.}(2012){Zavattini}, {Gastaldi}, {Pengo},
  {Ruoso}, {Della Valle}, and {Milotti}]{2012IJMPA..2760017Z}
{Zavattini}, G.; {Gastaldi}, U.; {Pengo}, R.; {Ruoso}, G.; {Della Valle}, F.;
  {Milotti}, E.
\newblock {Measuring the Magnetic Birefringence of Vacuum: the Pvlas
  Experiment}.
\newblock {\em International Journal of Modern Physics A} {\bf 2012}, {\em
  27},~1260017,  \href{http://xxx.lanl.gov/abs/1201.2309}{{\normalfont
  [arXiv:hep-ex/1201.2309]}}.

\bibitem[{Hartman} \em{et~al.}(2017){Hartman}, {Riv{\`e}re}, {Battesti}, and
  {Rizzo}]{2017RScI...88l3114H}
{Hartman}, M.T.; {Riv{\`e}re}, A.; {Battesti}, R.; {Rizzo}, C.
\newblock {Noise characterization for resonantly enhanced polarimetric vacuum
  magnetic-birefringence experiments}.
\newblock {\em Review of Scientific Instruments} {\bf 2017}, {\em 88},~123114,
  \href{http://xxx.lanl.gov/abs/1712.01278}{{\normalfont
  [arXiv:physics.ins-det/1712.01278]}}.

\bibitem[{King} and {Elkina}(2016)]{2016PhRvA..94f2102K}
{King}, B.; {Elkina}, N.
\newblock {Vacuum birefringence in high-energy laser-electron collisions}.
\newblock {\em \pra} {\bf 2016}, {\em 94},~062102,
  \href{http://xxx.lanl.gov/abs/1603.06946}{{\normalfont
  [arXiv:hep-ph/1603.06946]}}.

\bibitem[{Hill} and {Roso}(2017)]{2017JPhCS.869a2015H}
{Hill}, III, W.T.; {Roso}, L.
\newblock {Probing the quantum vacuum with petawatt lasers}.
\newblock  Journal of Physics Conference Series,  2017, Vol. 869, {\em Journal
  of Physics Conference Series}, p. 012015.

\bibitem[{Kii}(1987)]{1987PASJ...39..781K}
{Kii}, T.
\newblock {X-ray polarizations from accreting strongly magnetized neutron stars
  - Case studies for the X-ray pulsars 4U 1626-67 and Hercules X-1}.
\newblock {\em \pasj} {\bf 1987}, {\em 39},~781--800.

\bibitem[Heyl and Shaviv(2002)]{Heyl01qed}
Heyl, J.S.; Shaviv, N.J.
\newblock QED and the High Polarization of the Thermal Radiation from Neutron
  Stars.
\newblock {\em \prd} {\bf 2002}, {\em 66},~023002 (4 pages).

\bibitem[{Weisskopf} \em{et~al.}(2016){Weisskopf}, {Ramsey}, {O'Dell},
  {Tennant}, {Elsner}, {Soffitta}, {Bellazzini}, {Costa}, {Kolodziejczak},
  {Kaspi}, {Muleri}, {Marshall}, {Matt}, and {Romani}]{2016SPIE.9905E..17W}
{Weisskopf}, M.C.; {Ramsey}, B.; {O'Dell}, S.; {Tennant}, A.; {Elsner}, R.;
  {Soffitta}, P.; {Bellazzini}, R.; {Costa}, E.; {Kolodziejczak}, J.; {Kaspi},
  V.; {Muleri}, F.; {Marshall}, H.; {Matt}, G.; {Romani}, R.
\newblock {The Imaging X-ray Polarimetry Explorer (IXPE)}.
\newblock  Space Telescopes and Instrumentation 2016: Ultraviolet to Gamma Ray,
   2016, Vol. 9905, {\em \procspie}, p. 990517.

\bibitem[Zhang \em{et~al.}(2016)Zhang et~al.]{2016arXiv160708823Z}
Zhang, S.N.; others.
\newblock eXTP -- enhanced X-ray Timing and Polarimetry Mission.
\newblock {\em Proc. SPIE} {\bf 2016}, p. 99051Q,
  \href{http://xxx.lanl.gov/abs/1607.08823}{{\normalfont
  [arXiv:astro-ph.IM/1607.08823]}}.

\bibitem[{Soffitta} \em{et~al.}(2016){Soffitta} et~al.]{2016SPIE.9905E..15S}
{Soffitta}, P.; others.
\newblock {XIPE: the x-ray imaging polarimetry explorer}.
\newblock  Space Telescopes and Instrumentation 2016: Ultraviolet to Gamma Ray,
   2016, Vol. 9905, {\em \procspie}, p. 990515.

\bibitem[{Ho} and {Lai}(2003)]{2003MNRAS.338..233H}
{Ho}, W.C.G.; {Lai}, D.
\newblock {Atmospheres and spectra of strongly magnetized neutron stars - II.
  The effect of vacuum polarization}.
\newblock {\em \mnras} {\bf 2003}, {\em 338},~233--252,
  \href{http://xxx.lanl.gov/abs/astro-ph/0201380}{{\normalfont
  [astro-ph/0201380]}}.

\bibitem[Dittrich and Reuter(1985)]{Ditt85}
Dittrich, W.; Reuter, M.
\newblock {\em Effective Lagrangians in quantum electrodynamics};
  Springer-Verlag: Berlin,  1985.

\bibitem[Dittrich and Gies(2000)]{Ditt00}
Dittrich, W.; Gies, H.
\newblock {\em Probing the Quantum Vacuum}; Springer-Verlag: Berlin,  2000.

\bibitem[Itzykson and Zuber(1980)]{Itzy80}
Itzykson, C.; Zuber, J.B.
\newblock {\em Quantum Field Theory}; McGraw-Hill: New York,  1980.

\bibitem[Berestetskii \em{et~al.}(1982)Berestetskii, Lifshitz, and
  Pitaevskii]{Bere82}
Berestetskii, V.B.; Lifshitz, E.M.; Pitaevskii, L.P.
\newblock {\em Quantum Electrodynamics}, second ed.; Pergamon: Oxford,  1982.

\bibitem[Mandl and Shaw(1993)]{Mand93}
Mandl, F.; Shaw, G.
\newblock {\em Quantum Field Theory}, revised ed.; John Wiley \& Sons:
  Chichester,  1993.

\bibitem[Peskin and Schroeder(1995)]{Pesk95}
Peskin, M.E.; Schroeder, D.V.
\newblock {\em Introduction to Quantum Field Theory}; Addison-Wesley: Reading,
  Massachusetts,  1995.

\bibitem[{Gies} and {Langfeld}(2002)]{2002IJMPA..17..966G}
{Gies}, H.; {Langfeld}, K.
\newblock {Loops and Loop Clouds - A Numerical Approach to the Worldline
  Formalism in QED}.
\newblock {\em International Journal of Modern Physics A} {\bf 2002}, {\em
  17},~966--976,
  \href{http://xxx.lanl.gov/abs/arXiv:hep-ph/0112198}{{\normalfont
  [arXiv:hep-ph/0112198]}}.

\bibitem[Mazur and Heyl(2015)]{2014arXiv1407.7490M}
Mazur, D.; Heyl, J.S.
\newblock Casimir Interactions between Magnetic Flux Tubes in a Dense Lattice.
\newblock {\em \prd} {\bf 2015}, {\em 91},~065019,
  \href{http://xxx.lanl.gov/abs/1407.7490}{{\normalfont
  [arXiv:hep-th/1407.7490]}}.
\newblock (25 pages).

\bibitem[Adler(1971)]{Adle71}
Adler, S.L.
\newblock Photon Splitting and Photon Dispersion in a Strong Magnetic Field.
\newblock {\em Ann. Phys.} {\bf 1971}, {\em 67},~599.

\bibitem[Heyl and Hernquist(1997{\natexlab{a}})]{Heyl97index}
Heyl, J.S.; Hernquist, L.
\newblock The Birefringence and Dichroism of the QED Vacuum.
\newblock {\em \jpa} {\bf 1997}, {\em 30},~6485--6492.

\bibitem[Heyl and Hernquist(1997{\natexlab{b}})]{Heyl97hesplit}
Heyl, J.S.; Hernquist, L.
\newblock An Analytic Form for the Effective Lagrangian of QED and its
  Application to Pair Production and Photon Splitting.
\newblock {\em \prd} {\bf 1997}, {\em 55},~2449--2454.

\bibitem[{Kubo} and {Nagata}(1983)]{1983JOSA...73.1719K}
{Kubo}, H.; {Nagata}, R.
\newblock {Vector representation of behavior of polarized light in a weakly
  inhomogeneous medium with birefringence and dichroism}.
\newblock {\em Journal of the Optical Society of America (1917-1983)} {\bf
  1983}, {\em 73},~1719.

\bibitem[Heyl and Shaviv(2000)]{Heyl99polar}
Heyl, J.S.; Shaviv, N.J.
\newblock Polarization Evolution in Strong Magnetic Fields.
\newblock {\em \mn} {\bf 2000}, {\em 311},~555--564.

\bibitem[Shaviv \em{et~al.}(1999)Shaviv, Heyl, and Lithwick]{Shav98lens}
Shaviv, N.J.; Heyl, J.S.; Lithwick, Y.
\newblock Magnetic Lensing near Ultramagnetized Neutron Stars.
\newblock {\em \mn} {\bf 1999}, {\em 306},~333--347.

\bibitem[{Taverna} \em{et~al.}(2014){Taverna}, {Muleri}, {Turolla}, {Soffitta},
  {Fabiani}, and {Nobili}]{2014MNRAS.438.1686T}
{Taverna}, R.; {Muleri}, F.; {Turolla}, R.; {Soffitta}, P.; {Fabiani}, S.;
  {Nobili}, L.
\newblock {Probing magnetar magnetosphere through X-ray polarization
  measurements}.
\newblock {\em \mnras} {\bf 2014}, {\em 438},~1686--1697,
  \href{http://xxx.lanl.gov/abs/1311.7500}{{\normalfont
  [arXiv:astro-ph.HE/1311.7500]}}.

\bibitem[Caiazzo and Heyl(2018)]{Caia18bh}
Caiazzo, I.; Heyl, J.
\newblock {Vacuum birefringence and the x-ray polarization from black-hole
  accretion disks}.
\newblock {\em \prd} {\bf 2018}, {\em 97},~083001,
  \href{http://xxx.lanl.gov/abs/1803.03798}{{\normalfont
  [arXiv:astro-ph.HE/1803.03798]}}.

\bibitem[{Gaenther} \em{et~al.}(2017){Gaenther}, {Egan}, {Heilmann}, {Heine},
  {Hellickson}, {Frost}, {Schulz}, and {Theriault-Shay}]{SPIE_REDSoX}
{Gaenther}, H.M.; {Egan}, M.; {Heilmann}, R.K.; {Heine}, S.N.T.; {Hellickson},
  T.; {Frost}, J.and~{Marshall}, H.L.; {Schulz}, N.S.; {Theriault-Shay}, A.
\newblock {REDSoX: Monte-Carlo ray-tracing for a soft x-ray spectroscopy
  polarimeter}.
\newblock  2017, Vol. 10399, pp. 10399 -- 10399 -- 13.

\bibitem[{Tendulkar} \em{et~al.}(2015){Tendulkar}, {Hasc{\"o}et}, {Yang},
  {Kaspi}, {Beloborodov}, {An}, {Bachetti}, {Boggs}, {Christensen}, {Craig},
  {Guillot}, {Hailey}, {Harrison}, {Stern}, and {Zhang}]{2015ApJ...808...32T}
{Tendulkar}, S.P.; {Hasc{\"o}et}, R.; {Yang}, C.; {Kaspi}, V.M.; {Beloborodov},
  A.M.; {An}, H.; {Bachetti}, M.; {Boggs}, S.E.; {Christensen}, F.E.; {Craig},
  W.W.; {Guillot}, S.; {Hailey}, C.A.; {Harrison}, F.A.; {Stern}, D.; {Zhang},
  W.
\newblock {Phase-resolved NuSTAR and Swift-XRT Observations of Magnetar 4U
  0142+61}.
\newblock {\em \apj} {\bf 2015}, {\em 808},~32,
  \href{http://xxx.lanl.gov/abs/1506.03098}{{\normalfont
  [arXiv:astro-ph.HE/1506.03098]}}.

\bibitem[Heyl \em{et~al.}(2003)Heyl, Shaviv, and Lloyd]{Heyl01polar}
Heyl, J.S.; Shaviv, N.J.; Lloyd, D.
\newblock The High-Energy Polarization-Limiting Radius of Neutron Star
  Magnetospheres: I. Slowly Rotating Neutron Stars.
\newblock {\em \mn} {\bf 2003}, {\em 342},~134--144.

\bibitem[{Ho} \em{et~al.}(2003){Ho}, {Lai}, {Potekhin}, and
  {Chabrier}]{2003ApJ...599.1293H}
{Ho}, W.C.G.; {Lai}, D.; {Potekhin}, A.Y.; {Chabrier}, G.
\newblock {Atmospheres and Spectra of Strongly Magnetized Neutron Stars. III.
  Partially Ionized Hydrogen Models}.
\newblock {\em \apj} {\bf 2003}, {\em 599},~1293--1301,
  \href{http://xxx.lanl.gov/abs/astro-ph/0309261}{{\normalfont
  [astro-ph/0309261]}}.

\bibitem[{Heyl} \em{et~al.}(2005){Heyl}, {Lloyd}, and
  {Shaviv}]{2005astro.ph..2351H}
{Heyl}, J.S.; {Lloyd}, D.; {Shaviv}, N.J.
\newblock {The High-Energy Polarization-Limiting Radius of Neutron Star
  Magnetospheres II -- Magnetized Hydrogen Atmospheres}.
\newblock {\em ArXiv Astrophysics e-prints} {\bf 2005},
  \href{http://xxx.lanl.gov/abs/astro-ph/0502351}{{\normalfont
  [astro-ph/0502351]}}.

\bibitem[{van Adelsberg} and {Perna}(2009)]{2009MNRAS.399.1523V}
{van Adelsberg}, M.; {Perna}, R.
\newblock {Soft X-ray polarization in thermal magnetar emission}.
\newblock {\em \mnras} {\bf 2009}, {\em 399},~1523--1533,
  \href{http://xxx.lanl.gov/abs/0907.3499}{{\normalfont
  [arXiv:astro-ph.HE/0907.3499]}}.

\bibitem[{Potekhin} \em{et~al.}(2016){Potekhin}, {Ho}, and
  {Chabrier}]{2016arXiv160501281P}
{Potekhin}, A.Y.; {Ho}, W.C.G.; {Chabrier}, G.
\newblock {Atmospheres and radiating surfaces of neutron stars with strong
  magnetic fields}.
\newblock {\em ArXiv e-prints} {\bf 2016},
  \href{http://xxx.lanl.gov/abs/1605.01281}{{\normalfont
  [arXiv:astro-ph.HE/1605.01281]}}.

\bibitem[{Baldini} \em{et~al.}(2016){Baldini}, {Muleri}, {Soffitta}, {Omodei},
  {Pesce-Rollins}, {Sgro}, {Latronico}, {Spada}, {Manfreda}, and {Di
  Lalla}]{2016cosp...41E.129B}
{Baldini}, L.; {Muleri}, F.; {Soffitta}, P.; {Omodei}, N.; {Pesce-Rollins}, M.;
  {Sgro}, C.; {Latronico}, L.; {Spada}, F.; {Manfreda}, A.; {Di Lalla}, N.
\newblock {Ximpol: a new X-ray polarimetry observation-simulation and analysis
  framework}.
\newblock  41st COSPAR Scientific Assembly,  2016, Vol.~41, {\em COSPAR
  Meeting}.

\bibitem[{Meszaros} \em{et~al.}(1988){Meszaros}, {Novick}, {Szentgyorgyi},
  {Chanan}, and {Weisskopf}]{1988ApJ...324.1056M}
{Meszaros}, P.; {Novick}, R.; {Szentgyorgyi}, A.; {Chanan}, G.A.; {Weisskopf},
  M.C.
\newblock {Astrophysical implications and observational prospects of X-ray
  polarimetry}.
\newblock {\em \apj} {\bf 1988}, {\em 324},~1056--1067.

\bibitem[{Wang} and {Lai}(2009)]{2009MNRAS.398..515W}
{Wang}, C.; {Lai}, D.
\newblock {Polarization evolution in a strongly magnetized vacuum: QED effect
  and polarized X-ray emission from magnetized neutron stars}.
\newblock {\em \mnras} {\bf 2009}, {\em 398},~515--527,
  \href{http://xxx.lanl.gov/abs/0903.2094}{{\normalfont
  [arXiv:astro-ph.HE/0903.2094]}}.

\end{thebibliography}

%

\end{document}